\documentclass[preprint2]{proto}
\usepackage{natbib}
\usepackage{pdfpages}
\bibpunct{(}{)}{;}{a}{,}{,}
\usepackage{times}
\usepackage[section]{placeins}
\usepackage{fixltx2e}

\newcommand{\mearth}{$M_\earth$}
\newcommand{\rearth}{$R_\earth$}
\newcommand{\msini}{$M \sin i$}
\usepackage{amsmath} 

\begin{document}

\title{\textbf{\LARGE Exoplanet Detection Techniques}}


 
\author {Debra A. Fischer\altaffilmark{1},
	     Andrew W.\ Howard\altaffilmark{2}, 
	     Greg P. Laughlin\altaffilmark{3},
	     Bruce Macintosh\altaffilmark{4},
	     Suvrath Mahadevan\altaffilmark{5,6},
              Johannes Sahlmann\altaffilmark{7},  
              Jennifer C. Yee\altaffilmark{8}}

\altaffiltext{1}{Department of Astronomy, Yale University, New Haven, CT 06520, USA}
\altaffiltext{2}{Institute for Astronomy, University of Hawai`i at Manoa, 
2680 Woodlawn Drive, Honolulu, HI
96822, USA}
\altaffiltext{3}{UCO/Lick Observatory, University of California at Santa Cruz, 
Santa Cruz, CA 95064, USA}
\altaffiltext{4}{Stanford University, Palo Alto CA USA}
\altaffiltext{5}{Center for exoplanets and Habitable Worlds, The Pennsylvania State University, University Park, PA 16802, USA} 
\altaffiltext{6}{Department of Astronomy and Astrophysics, The Pennsylvania State University, University Park, PA 16802, USA} 
\altaffiltext{7}{Observatoire de Gen\`eve, Universit\'e de Gen\`eve, 51 Chemin 
Des Maillettes, 1290 Versoix, Switzerland}
\altaffiltext{8}{Harvard-Smithsonian Center for Astrophysics, 60 Garden St, Cambridge, MA 02138 USA}  

\begin{abstract}
We are still in the early days of exoplanet discovery. Astronomers are
beginning to model the atmospheres and interiors of exoplanets and have
developed a deeper understanding of processes of planet formation and
evolution. However, we have yet to map out the full complexity of
multi-planet architectures or to detect Earth analogues around nearby
stars. Reaching these ambitious goals will require further improvements
in instrumentation and new analysis tools. In this chapter, we provide
an overview of five observational techniques that are currently employed
in the detection of exoplanets: optical and IR Doppler measurements,
transit photometry, direct imaging, microlensing, and astrometry. We
provide a basic description of how each of these techniques works and
discuss forefront developments that will result in new discoveries. We
also highlight the observational limitations and synergies of each
method and their connections to future space missions.

\end{abstract} 


\section{Introduction} 
Humans have long wondered whether other solar systems 
exist around the billions of stars in our galaxy.  In the past two decades, we 
have progressed from a sample of one to a collection of hundreds of exoplanetary systems. 
Instead of an orderly solar nebula model, we now realize that chaos rules the formation  
of planetary systems.  Gas giant planets can migrate close to their stars. Small rocky 
planets are abundant and dynamically pack the inner orbits.  Planets circle outside the 
orbits of binary star systems. The diversity is astonishing. 

Several methods for detecting exoplanets have been developed: Doppler measurements,
transit observations, microlensing, astrometry, and direct imaging.  Clever 
innovations have advanced the precision for each of these techniques, however 
each of the methods have inherent observational incompleteness. The lens through 
which we detect exoplanetary systems biases the parameter space that we can 
see. For example, Doppler and transit techniques preferentially detect 
planets that orbit closer to their host stars and are larger in mass or size while 
microlensing, astrometry, and direct imaging are more sensitive to planets in 
wider orbits. In principle, the techniques are complementary; in practice, they are not 
generally applied to the same sample of stars, so our detection of exoplanet 
architectures has been piecemeal. The explored parameter space of exoplanet systems 
is a patchwork quilt that still has several missing squares.

\section{The Doppler Technique}

\subsection{Historical Perspective}
The first Doppler detected planets were met with skepticism. \citet{CWY88} identified 
variations in the residual velocities of $\gamma$ Ceph, a component of a binary star system, 
but attributed them to stellar activity signals until additional data confirmed this as a 
planet fifteen years later \citep{H03}.  \citet{L89} detected a Doppler signal around HD~114762 with an 
orbital period of 84 days and a mass $M_P \sin i = 11 M_{Jup}$. Since the orbital inclination was 
unknown, they expected that the mass could be significantly larger and 
interpreted their data as a probable brown dwarf.   When \citet{MQ95} modeled a 
Doppler signal in their data for the sunlike star, 51 Pegasi, as a Jupiter-mass planet 
in a 4.23-day orbit, astronomers wondered if this could be a previously unknown 
mode of stellar oscillations \citep{G97} or non-radial pulsations \citep{H97}. 
The unexpected detection of significant eccentricity in 
exoplanet candidates further raised doubts among astronomers who argued that 
although stars existed in eccentric orbits, planets should reside in circular orbits 
\citep{B97}. It was not until the first transiting planet \citep{H00, C00} and the first 
multi-planet system \citep{B99} were detected (almost back-to-back) that the planet 
interpretation of the Doppler velocity data was almost unanimously accepted. 

The Doppler precision improved from about 10 ${\rm m\, s^{-1}}$ in 1995 to 
3 ${\rm m\, s^{-1}}$ in 1998, and then to about 1 ${\rm m\, s^{-1}}$ in 2005 when HARPS was 
commissioned \citep{M03}.  A Doppler precision 
of 1 ${\rm m \,s^{-1}}$ corresponds to shifts of stellar lines across 1/1000th of a CCD pixel.  
This is a challenging measurement that requires high signal-to-noise,  high-resolution, 
and large spectral coverage. Echelle spectrometers typically provide these attributes 
and have served as the workhorse instruments for Doppler planet searches.

\begin{figure}[tb]
\begin{center}
\includegraphics[width=1.\columnwidth]{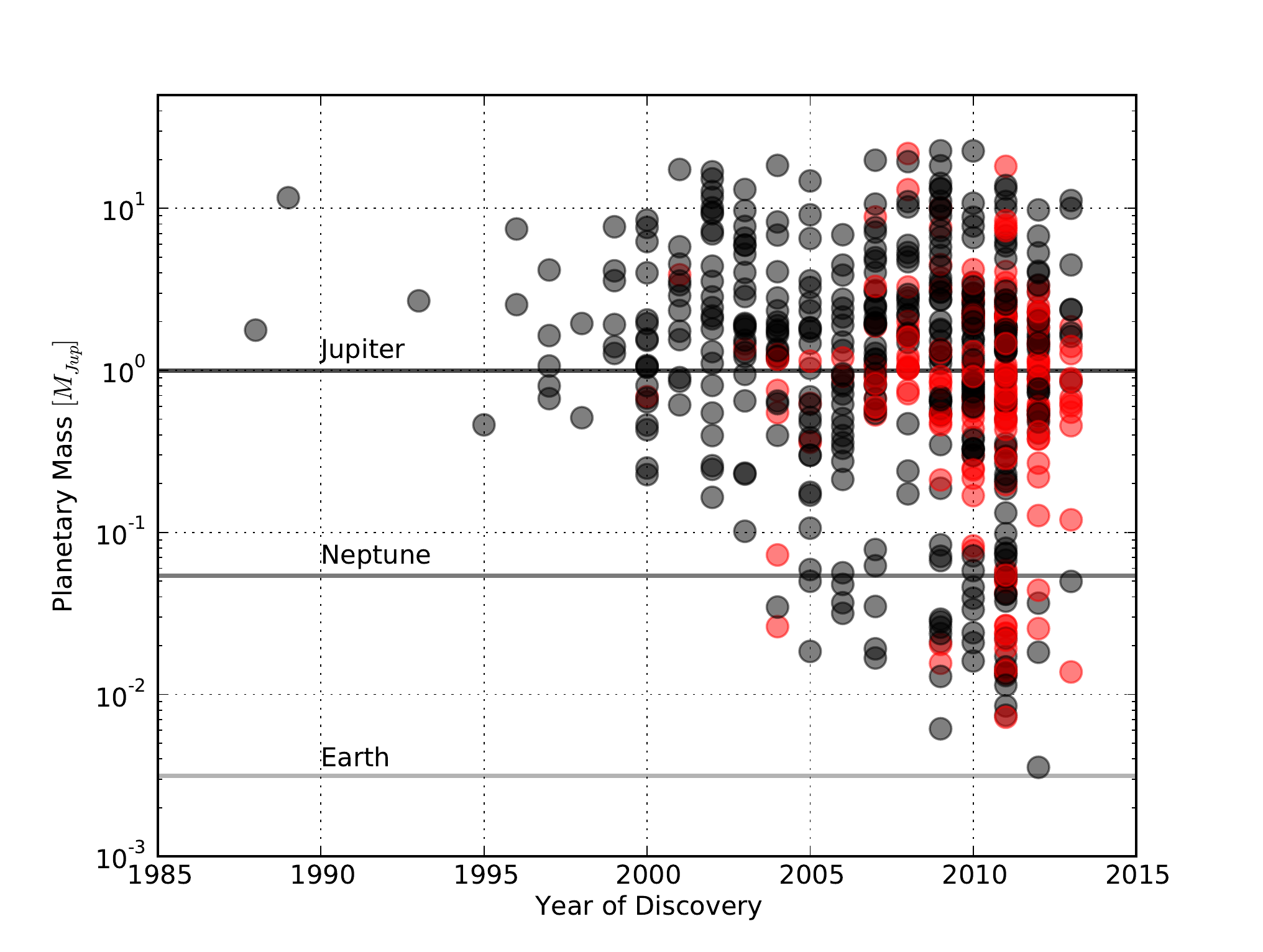}
\caption{\label{fig:yr-mass} Planet mass is plotted as a function of the year of discovery. 
The color coding is gray for planets with no known transit, whereas light red is planets 
that do transit.}
\end{center}
\end{figure}

Figure \ref{fig:yr-mass} shows the detection history for planets identified with 
Doppler surveys (planets that also are observed to transit their host star are 
color-coded in red). The first planets were similar in mass to Jupiter and there has been
a striking decline in the lower envelope of detected planet mass with time as 
instrumentation improved.

\subsection{Radial Velocity Measurements}
The Doppler technique measures the reflex velocity that an orbiting planet 
induces on a star. Because the star-planet interaction is mediated by gravity, more 
massive planets result in larger and more easily detected stellar velocity amplitudes. 
It is also easier to detect close-in planets, both because the gravitational
force increases with the square of the distance and because the orbital periods are 
shorter and therefore more quickly detected. \citet{LF10} provide a detailed discussion 
of the technical aspects of Doppler analysis with both an iodine cell and a thorium-argon 
simultaneous reference source. 

The radial velocity semi-amplitude, $K_1$ of the star can be expressed in units of ${\rm cm\,s^{-1}}$ with the planet 
mass in units of $M_\oplus$:

\begin{equation}
\label{eqn:K}
K_* \!= \! \frac{8.95\,\mathrm{cm\,s^{-1}}}{\sqrt{1\!-\!e^2}} \, \frac{M_P \sin{i}}{M_{\oplus}} 
\! \left( \frac{M_*\!+\!M_P}{M_\odot}\right)^{-2/3} \left(\frac{P}{\mathrm{yr}} \right)^{-1/3} 
\end{equation}

The observed parameters (velocity semi-amplitude $K_*$, orbital period $P$, eccentricity $e$,
and orientation angle $\omega$)
are used to calculate a minimum mass of the planet $M_P \sin i$ if the mass of the 
star $M_*$ is known.  The true mass of the planet is unknown because
it is modulated by the unknown inclination. For example, if the orbital inclination 
is thirty degrees, the true mass is a factor of two times the Doppler-derived 
$M_P \sin i$.  The statistical probability that the orbit inclination
is within an arbitrary range $i_1 < i < i_2$ is given by 
\begin{equation} 
\label{eqn:prob_i}
{\cal P}_{incl} = | \cos (i_2) - \cos (i_1) |
\end{equation}

Thus, there is a roughly 87\% probability that random orbital inclinations 
are between thirty and ninety degrees, or equivalently, an 87\% probability that the 
true mass is within a factor of two of the minimum mass $M_P \sin i$.

Radial velocity observations must cover one complete orbit in order
to robustly measure the orbital period. As a result the first 
detected exoplanets resided in short-period orbits. Doppler surveys that 
have continued for a decade or more \citep{Fischer13, Marmier13} have 
been able to detect gas giant planets in Jupiter-like orbits.

\subsection{The floor of the Doppler precision}
An important question is whether the Doppler technique can be further improved 
to detect smaller planets at wider orbital 
radii. The number of exoplanets detected each year rose steadily until 2011 
and has dropped precipitously after that year. This is due in part to the fact that 
significant telescope time has been dedicated to transit follow-up and also because 
observers are working to extract the smallest possible planets, requiring more 
Doppler measurement points given current precision. 
Further gains in Doppler precision and productivity will require new instruments with greater stability 
as well as analytical techniques for decorrelating stellar noise.  

\begin{figure}[tb]
\begin{center}
\includegraphics[width=\columnwidth]{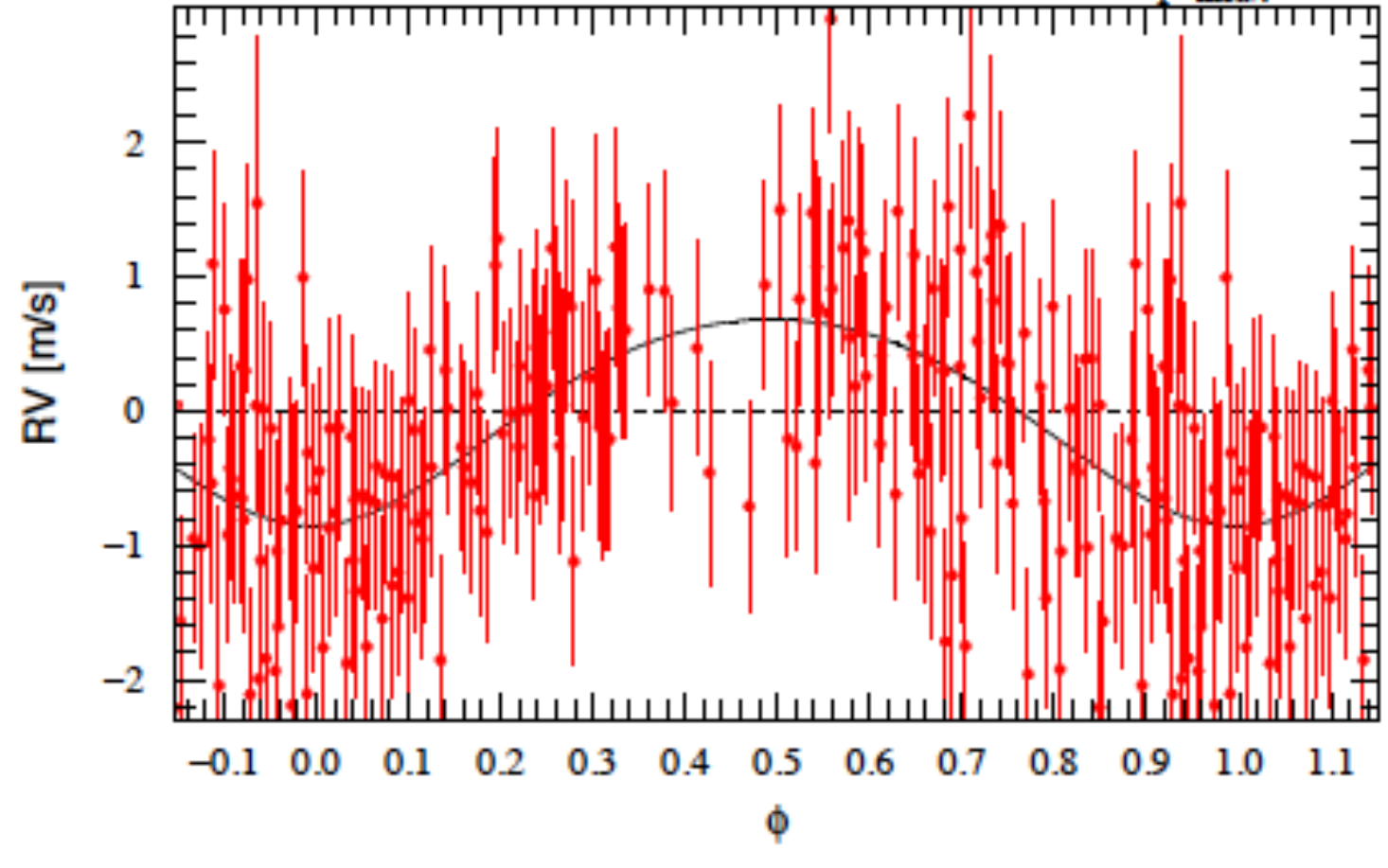}
\caption{\label{fig:HD85512} The phase-folded data for the detection of a planet orbiting 
HD~85512 (Figure 13 from Pepe et al. 2011). }
\end{center}
\end{figure}

Figure \ref{fig:HD85512}, reproduced from \citet{P11}, 
shows an example of one of the lowest amplitude exoplanets, detected with HARPS. 
The velocity semi-amplitude for this planet is $K = 0.769\, {\rm m\, s^{-1}}$ and the 
orbital period is 58.43 days. The data was comprised of 185 observations spanning 7.5 years. 
The residual velocity scatter after fitting for the planet was reported to 
be $0.77 \,{\rm m\, s^{-1}}$, showing that high precision can be achieved with 
many data points to beat down the single measurement precision. 

One promising result suggests that it may be possible for stable spectrometers to average 
over stellar noise signals and reach precisions below 0.5 ${\rm m\, s^{-1}}$, 
at least for some stars.  
After fitting for three planets in HD20794, \citet{P11} found that the 
RMS of the residual velocities decreased from 0.8 ${\rm m\, s^{-1}}$ to 
0.2 ${\rm m\, s^{-1}}$ as they binned the data in intervals from 1 to 40 nights.
Indeed, a year later, the HARPS team published the smallest velocity 
signal ever detected: a planet candidate 
that orbits alpha Centauri B \citep{Dumusque12} with a 
velocity amplitude  $K = 0.51\, {\rm m\, s^{-1}}$, planet 
mass $M \sin i = 1.13 \,M_{\oplus}$, and an orbital period of 3.24 days.  
This detection required 469 Doppler measurements obtained over 7 years 
and fit for several time-variable stellar noise signals.
Thus, the number of observations required to solve for the 5-parameter Keplerian
model increases exponentially with decreasing velocity amplitude.  

\subsection{The Future of Doppler Detections}
It is worth pondering whether improved instruments with higher resolution, higher sampling,
greater stability and more precise wavelength calibration will ultimately be 
able to detect analogs of the Earth with 0.1\, ${\rm m\, s^{-1}}$ velocity amplitudes.
An extreme precision spectrometer will have stringent environmental requirements 
to control temperature, pressure and vibrations. The dual requirements of 
high resolution and high signal-to-noise lead to the need for moderate
to large aperture telescopes \citep{S08, S12}. The coupling of light into the instrument must be 
exquisitely stable.  This can be achieved with a double fiber scrambler \citep{H92} 
where the near field of the input fiber is mapped to the far field of the output fiber, 
providing a high level of scrambling in both the radial and azimuthal directions. At some cost to 
throughput, the double fiber scrambler stabilizes variations in the 
spectral line spread function (sometimes called a point spread function) 
and produces a series of spectra that are uniform except for photon noise. 
Although the fibers provide superior illumination of the spectrometer optics, 
some additional care in the instrument design phase is required to provide excellent flat fielding and 
sky subtraction.  The list of challenges to extreme instrumental precision also 
includes the optical CCD detectors, 
with intra-pixel quantum efficiency variations, tiny variations in pixel sizes, 
charge diffusion and the need for precise controller software to perfectly 
clock readout of the detector. 

In addition to the instrumental precision, another challenge to high Doppler precision 
is the star itself. 
Stellar activity, including star spots, p-mode oscillations and variable granulation are 
tied to changes in the strength of stellar magnetic fields. These stellar 
noise sources are sometimes called stellar jitter and can produce 
line profile variations that skew the center of mass for a spectral line in a 
way that is (mis)interpreted by a Doppler code as a velocity change in the 
star. Although stellar noise signals are subtle, they affect the spectrum in a 
different way than dynamical velocities. 
The stellar noise typically has a color dependence and an asymmetric velocity 
component. in order to reach significantly higher accuracy in velocity measurements, 
it is likely that we will need to identify and model or decorrelate 
the stellar noise. 

\section{Infrared Spectroscopy} 
\subsection{Doppler Radial Velocities in the Near Infrared}
The high fraction of Earth-size planets estimated to orbit in the habitable zones (HZs) of M dwarfs \citep{Dressing2013, Kopparapu2013, Bonfils2013} makes the low mass stars very attractive targets for Doppler RV surveys.  The lower stellar mass of the M dwarfs, as well as the short orbital periods of HZ planets, increases the amplitude of the Doppler wobble (and the ease of its detectability) caused by such a terrestrial-mass planet. However, nearly all the stars in current optical RV surveys are earlier in spectral type than $\sim$M5 since later spectral types are difficult targets even on large telescopes due to their intrinsic faintness in the optical: they emit most of their flux in the red optical and near infrared (NIR) between 0.8 and 1.8 $\mu$m (the NIR Y, J and H bands are 0.98-1.1 $\mu$m, 1.1-1.4 $\mu$m and 1.45-1.8 $\mu$m). However, it is the low mass late-type M stars, which are the least luminous, where the velocity amplitude of a terrestrial planet in the habitable zone is highest, making them very desirable targets. Since the flux distribution from M stars peaks sharply in the NIR, stable high-resolution NIR spectrographs capable of delivering high RV precision can observe several hundred of the nearest M dwarfs to examine their planet population. 

\subsubsection{Fiber-Fed NIR High-Resolution Spectrographs}
A number of new fiber-fed stabilized spectrographs are now being designed and built for such a purpose: the Habitable Zone Planet finder \citep{Mahadevan12} for the 10m Hobby Eberly Telescope, CARMENES \citep{Quirrenbach12} for the 3.6m Calar Alto Telescope and Spirou \citep{Santerne13} being considered for the CFHT. The instrumental challenges in the NIR, compared to the optical, are calibration, stable cold operating temperatures of the instrument, and the need to use NIR detectors. The calibration issues seem tractable (see below).  Detection of light beyond 1$\mu m$ required the use of NIR sensitive detectors like the Hawaii-2(or 4)RG HgCdTe detectors. These devices are fundamentally different than CCDs and exhibit effects like inter-pixel capacitance and much greater persistence. Initial concerns about the ability to perform precision RV measurements with these device has largely been retired with lab \citep{Ramsey08} and on sky demonstrations \citep{Ycas12a} with a Pathfinder spectrograph, though careful attention to ameliorating these effects is still necessary to achieve high RV precision. This upcoming generation of spectrographs, being built to deliver 1-3 ${\rm m\,s^{-1}}$ RV precision in the NIR will also be able to confirm many of the planets detected with TESS and Gaia around low mass stars. NIR spectroscopy is also a essential tool to be able to discriminate between giant planets and stellar activity in the search for planets around young active stars \citep{Mahmud11}.

\subsubsection{Calibration Sources}
Unlike iodine in the optical no single known gas cell simultaneously covers large parts of the NIR z, Y, J \& H bands. Thorium Argon lamps, that are so successfully used in the optical have very few Thorium emission lines in the NIR, making them unsuitable as the calibrator of choice in this wavelength regime. Uranium has been shown to provide a significant increase in the number of lines available for precision wavelength calibration in the NIR. New linelists have been published for Uranium lamps \citep{Redman11a, Redman11b} and these lamps are now in use in existing and newly commissioned NIR spectrographs.  Laser frequency combs, which offer the prospects of very high precision and accuracy in wavelength calibration, have also been demonstrated with astronomical spectrographs in the NIR \citep{Ycas12a} with filtering making them suitable for an astronomical spectrograph. Generation of combs spanning the entire z-H band regions has also been demonstrated in the lab \citep{Ycas12b}. Continuing development efforts are aimed at effectively integrating these combs as calibration sources for M dwarf Doppler surveys with stabilized NIR spectrographs.  Single mode fiber-based Fabry-P{\'e}rot cavities fed by supercontinuum light sources have also been demonstrated by \citet{Halverson12}. To most astronomical spectrographs the output from these devices looks similar to that of a laser comb, although the frequency of the emission peaks is not known innately to high precision. Such inexpensive and rugged devices may soon be available for most NIR spectrographs, with the superior (and more expensive) laser combs being reserved for the most stable instruments on the larger facilities. While much work remains to be done to refine these calibration sources, the calibration issues in the NIR largely seem to be within reach.

\subsubsection{Single Mode Fiber-fed Spectrographs}
The advent of high strehl ratio adaptive optics (AO) systems at most large telescopes makes it possible to seriously consider using a single-mode optical fiber (SMF) to couple the light from the focal plane of the telescope to a spectrograph. Working close to the diffraction limit enables such SMF-fed spectrographs to be very compact while simultaneously capable of providing spectral resolution comparable or superior to natural seeing spectrographs.  A number of groups are pursuing technology development relating to these goals \citep{Ghasempour12, Schwab13, Crepp13}. The single mode fibers provide theoretically perfect scrambling of the input PSF, further aiding in the possibility of very high precision and compact Doppler spectrometers emerging from such development paths. While subtleties relating to polarization state and its impact on velocity precision remain to be solved, many of the calibration sources discussed above are innately adaptable to use with SMF fiber-fed spectrographs.  Since the efficiency of these systems depends steeply on the level of AO correction, it is likely that Doppler RV searches targeting the red optical and NIR wavelengths will benefit the most.

\subsection{Spectroscopic Detection of Planetary Companions}
Direct spectroscopic detection of the orbit of non-transiting planets has finally yielded successful results this decade. While the traditional Doppler technique relies of detecting the radial velocity of the star only,  the direct spectroscopic detection technique relies on observing the star-planet system in the NIR or thermal IR (where the planet to star flux ratio is more favorable than the optical) and obtaining high resolution, very high S/N spectra to be able to spectroscopically measure the radial velocity of both the star {\it and the planet}  in a manner analogous to the detection of a spectroscopic binary (SB2). The radial velocity observations directly yield the mass ratio of the star-planet system. If the stellar mass is known (or estimated well) the planet mass can be determined with no $\sin{i}$ ambiguity despite the fact that these are not transiting systems. The spectroscopic signature of planets orbiting Tau Boo, 51 Peg, and HD189733 have recently been detected using the CRIRES instrument on the VLT \citep{Brogi12, Brogi13, Kok13, Rodler12} and efforts are ongoing by multiple groups to detect other systems using the NIRSPEC instrument at Keck \citep{Lockwood2014}.  The very high S/N required of this technique limits it to the brighter planet hosts, and to rleatively close-in planets, but yields information about mass and planetary atmospheres that would be difficult to determine otherwise for the non-transiting planets. Such techniques complement the transit detection efforts underway and will increase in sensitivity with telescope aperture , better infrared detectors, and more sophisticated analysis techniques. While we have focused primarily on planet detection techniques in this review article, high resolution NIR spectroscopy using large future gound based telescopes may also be able to detect astrobiologically interesting molecules (eg. $O_2$) around Earth-analogues orbiting M dwarfs \citep{Snellen13}.

\section{Doppler Measurements from Space}
Although there are no current plans to build high-resolution spectrometers for space missions, 
this environment might offer some advantages for extreme precision Doppler spectroscopy if
the instrument would be in a stable thermal and pressure environment. Without blurring from the 
Earth's atmosphere, the point spread function (PSF) would be very stable and 
the image size could be small making it intrinsically easier to obtain high 
resolution with an extremely compact instrument. Furthermore, the effect of sky 
subtraction and telluric contamination are currently difficult problems to solve 
with ground-based instruments and these issues are eliminated with space-based 
instruments. 

\section{Transit Detections} 

At the time of the press run for the Protostars and Planets IV in 2000, the first 
transiting extrasolar planet -- HD 209458b -- had just been found  \citep{H00, C00}. 
That momentous announcement, however, was too late for the conference volume, and 
PPIV's single chapter on planet detection was devoted to fourteen planets detected 
by Doppler velocity monitoring, of which only eight were known prior to the June 1998 
meeting. Progress, however, was rapid. In 2007, when the Protostars and Planets V volume 
was published, nearly 200 planets had been found with Doppler radial velocities, and 
nine transiting planets were then known \citep{Charbonneau07}. 

In the past several years, the field of transit detection has come dramatically into 
its own. A number of long-running ground-based projects, notably the 
SuperWASP \citep{CollierCameron07} and HATNet surveys \citep{Bakos07}, have amassed 
the discovery of dozens of transiting planets with high-quality light curves in concert 
with accurate masses determined via precision Doppler velocity measurements. Thousands 
of additional transiting planetary candidates have been observed from space. Transit 
timing variations \citep{Agol05, Holman05} have progressed from a theoretical 
exercise to a practiced technique. The Spitzer Space Telescope (along with HST 
and ground-based assets) has been employed to characterize the atmospheres of 
dozens of transiting extrasolar planets \citep{Seager10}. An entirely new, and 
astonishingly populous, class of transiting planets in the mass 
range $R_{\oplus}<R_P<4R_{\oplus}$ has been discovered and probed \citep{Batalha13}. 
Certainly, with each new iteration of the Protostars and Planets series, the previous 
edition looks hopelessly quaint and out of date. Is seems certain that progress will 
ensure that this continues to be the case.

\subsection{The Era of Space-based Transit Discovery}

Two space missions, Kepler \citep{Borucki10} and CoRoT \citep{Barge08} have both 
exhibited excellent productivity, and a third mission, MOST, has provided photometric 
transit discoveries of several previously known planets \citep{Winn11, Dragomir13}. 
Indeed, Figure \ref{fig:yr-mass} indicates that during the past six years, transiting 
planets have come to dominate the roster of new discoveries. Doppler velocimetry, which 
was overwhelmingly the most productive discovery method through 2006, is rapidly 
transitioning from a general survey mode to an intensive focus on low-mass planets 
orbiting very nearby stars \citep{Mayor2011} and to the characterization of planets 
discovered in transit via photometry.

\begin{figure}[tb]
\includegraphics[width=1.\linewidth]{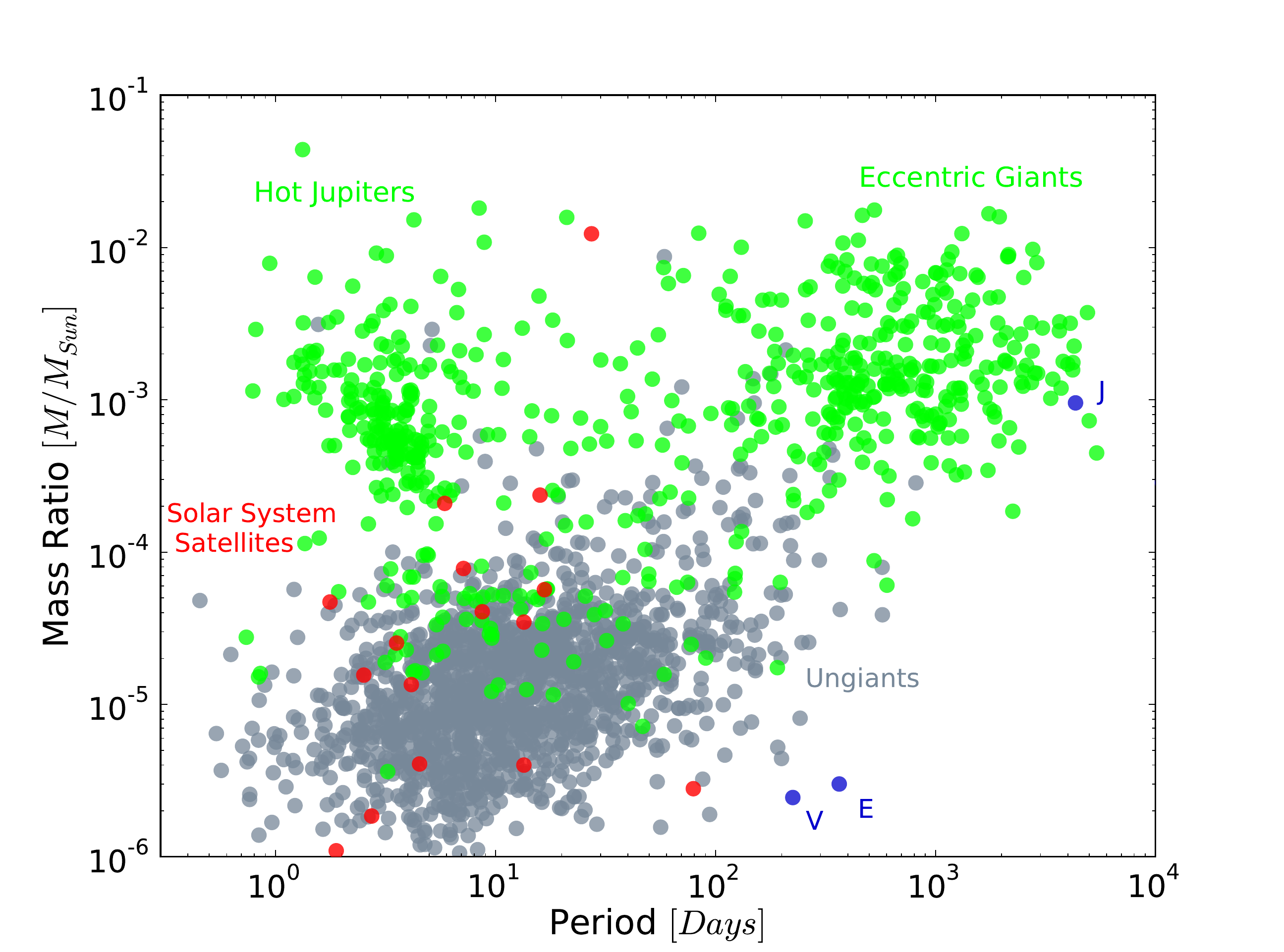}
\caption{{\it Green circles:} $\log_{10}(M_{\rm satellite}/M_{\rm primary})$ 
and $\log_{10}(P)$ for 634 planets securely detected by the radial velocity method 
(either with or without photometric transits). {\it Red circles:} 
$\log_{10}(M_{\rm satellite}/M_{\rm primary})$ and $\log_{10}(P)$ for the regular 
satellites of the Jovian planets in the Solar System.  {\it Gray circles:} 
$\log_{10}(M_{\rm satellite}/M_{\rm primary})$ and $\log_{10}(P)$ for 1501 Kepler 
candidates and objects of interest in which {\it multiple} transiting candidate 
planets are associated with a single primary. Radii for these candidate planets, 
as reported in \citep{Batalha13}, are converted to masses assuming 
$M/M_{\oplus}=(R/R_{\oplus})^{2.06}$ \citep{Lissauer2011}, which is obtained by fitting 
the masses and radii of the solar system planets bounded in mass by Venus and Saturn. 
Data are from www.exoplanets.org, accessed 08/15/2013. \label{fig:kepler-multis}}
\end{figure}

The Kepler Mission, in particular, has been completely transformative, having generated, 
at last rapidly evolving count, over one hundred planets with mass determinations, as well as hundreds 
of examples of multiple transiting planets orbiting a single host star, 
many of which are in highly co-planar, surprisingly crowded systems \citep{Lissauer2011b}. 
Taken in aggregate, the Kepler candidates indicate that planets with
masses $M_P<30M_{\oplus}$ and orbital periods, $P<100\,{\rm d}$ are effectively 
ubiquitous \citep{Batalha2011}, and as shown in Figure \ref{fig:kepler-multis}, the distribution of mass ratios and periods 
of these candidate planets are, in many cases, curiously reminiscent of the regular  satellites of the Jovian planets within our own solar system.

The CoRoT satellite ceased active data gathering in late 2012, having substantially 
exceeded its three-year design life. In Spring of 2013, just after the end of its 
nominal mission period, the Kepler satellite experienced a failure of a second 
reaction wheel, which brought its high-precision photometric monitoring program to 
a premature halt. The four years of Kepler data in hand, however, are well-curated, 
fully public, and are still far from being fully exploited; it is not unreasonable 
to expect that they will yield additional insight that is equivalent to what has 
already been gained from the mission to date.  \citet{Jenkins10} describe the fiducial 
Kepler pipeline; steady improvements to the analysis procedures therein have led to 
large successive increases in the number of planet candidates detected per 
star \citep{Batalha13}.

The loss of  the Kepler and CoRoT spacecraft has been tempered by the recent approvals 
of two new space missions. In the spring of 2013, NASA announced selection of the 
Transiting Exoplanet Survey Satellite (TESS) Mission for its Small Explorer Program. 
TESS is currently scheduled for a 2017 launch. It will employ an all-sky strategy to 
locate transiting planets with periods of weeks to months, and sizes down to 
$R_{\rm p}\sim1R_{\oplus}$ (for small parent stars) among a sample of $5\times10^{5}$ stars brighter 
than $V=12$, including $\sim1000$ red dwarfs \citep{Ricker10}. TESS is designed 
to take advantage of the fact that the most heavily studied, and therefore the most 
scientifically valuable, transiting planets in a given category (hot Jupiters, 
extremely inflated planets, sub-Neptune sized planets, etc.) orbit the brightest 
available parent stars. To date, many of these ``fiducial'' worlds, such as HD 209458~b
HD 149026~b, HD 189733~b,  and Gliese 436~b, have been discovered to transit by 
photometrically monitoring known Doppler-wobble stars during the time windows 
when transits are predicted to occur. By surveying {\it all} the bright stars, 
TESS will systematize the discovery of the optimal transiting example planets within 
every physical category. The CHEOPS satellite is also scheduled for launch in 2017 \citep{Broeg13}. It will 
complement TESS by selectively and intensively searching for transits by candidate 
planets in the $R_{\oplus}<R_{\rm p}<4R_{\oplus}$ size range during time windows 
that have been identified by high-precision Doppler monitoring of the parent stars. It will also perform follow-up observations of
interesting TESS candidates.

\subsection{Transit Detection}

The {\it a-priori} probability that a given planet can be observed in transit is 
a function of the planetary orbit, and the planetary and stellar radii
\begin{equation}
\label{eqn:transProb}
{\cal P}_{tr}=0.0045\left(\frac{{\rm AU}}{a}\! \right)\left(\frac{R_{\star}\!+\!R_{\rm p}}
{R_{\odot}}\right)\! \left[\frac{1\!+\!e\cos(\pi/2\!-\!\omega)}{1-e^2}\right]\, ,
\end{equation}
where $\omega$ is the angle at which orbital periastron occurs, such that 
$\omega=90^{\circ}$ indicates transit, and $e$ is the orbital eccentricity. 
A typical hot Jupiter with $R_{\rm p}\gtrsim R_{\rm Jup}$ and $P\sim3\,{\rm d}$, 
orbiting a solar-type star, has a $\tau \sim 3\,{\rm hr}$ transit duration, a 
photometric transit depth, $d\sim1$\%, and ${\cal P}\sim10$\%. Planets belonging 
to the ubiquitous super-Earth -- sub-Neptune population identified by Kepler (i.e., 
the gray points in Figure \ref{fig:kepler-multis}) are typified by 
${\cal P}\sim2.5$\%, $d\sim0.1$\%, and $\tau \sim 6\,{\rm hr}$, whereas 
Earth-sized planets in an Earth-like orbits around a solar-type stars present a 
challenging combination of ${\cal P}\sim0.5$\%, $d\sim0.01$\%, and 
$\tau \sim 15\,{\rm hr}$.

Effective transit search strategies seek the optimal trade-off between cost, 
sky coverage, photometric precision, and the median apparent brightness of the 
stars under observation. For nearly a decade, the community as a whole struggled 
to implement genuinely productive surveys. For an interesting summary of the 
early disconnect between expectations and reality, see \citet{Horne03}. Starting 
in the mid-2000s, however, a number of projects began to produce transiting 
planets \citep{Konacki03, Alonso04, McCollough06}, and there are now a range 
of successful operating surveys.  For example, the ongoing Kelt-North project, 
which has discovered 4 planets to date \citep{Collins13} targets very bright 
$8<V<10$ stars throughout a set of $26^{\circ}\times26^{\circ}$ fields that 
comprise $\sim$12\% of the full sky. Among nearly 50,000 stars in this survey, 
3,822 targets have RMS photometric precision better than 1\% (for 150-sec 
exposures). A large majority of the known transit-bearing stars, however, 
are fainter than Kelt's faint limit near $V\sim10$. The $10<V<12$ regime has 
been repeatedly demonstrated to provide good prospects for Doppler follow-up 
and detailed physical characterization, along with a large number of actual 
transiting planets. In this stellar brightness regime, surveys such as HATNet 
and SuperWASP have led the way. For instance, HAT-South \citep{Bakos13}, a 
globally networked extension of the long-running HATNet project, monitors 
$8.2^{\circ}\times8.2^{\circ}$ fields and reaches 6 millimagnitude (mmag) 
photometric precision at 4-minute cadence for the brightest non-saturated 
stars at $r\sim10.5$. SuperWASP's characteristics are roughly similar, and to date, it has been the most productive ground-based transit search program.

To date, the highest-precision ground-based exoplanetary photometry has been 
obtained with orthogonal phase transfer arrays trained on single, carefully 
preselected high-value target stars. Using this technique, \citep{Johnson09} 
obtained 0.47\,mmag photometry at 80-second cadency for WASP-10 (V=12.7). By 
comparison, with its space-borne vantage, Kepler obtained a median photometric 
precision of 29 ppm with 6.5 hour cadence on V=12 stars. This is $\sim2\times$ 
better than the best special-purpose ground-based photometry, and $\sim20\times$ 
better than the leading ground-based discovery surveys.

Astrophysical false positives present a serious challenge for wide-field surveys 
in general and for Kepler in particular, where a majority of the candidate planets 
lie effectively out of reach of Doppler characterization and confirmation 
\citep{Morton11}. Stars at the bottom of the main sequence overlap in size with 
giant planets \citep{Chabrier00} and thus present near-identical transit signatures 
to those of giant planets. Grazing eclipsing binaries can also provide a source of 
significant confusion for low signal-to-noise light curves \citep{Konacki03}.

Within the Kepler field, pixel ``blends'' constitute a major channel for false alarms. 
These occur when an eclipsing binary, either physically related or unrelated, shares 
line of sight with the target star. Photometry alone can be used to identify many 
such occurrences \citep{Batalha10}, whereas in other cases, statistical modeling of 
the likelihood of blend scenarios 
\citep{Torres04, Fressin2013} can establish convincingly low false alarm 
probabilities.  High-profile examples of confirmation by statistical validation 
include the $R=2.2R_{\oplus}$ terrestrial candidate planet Kepler 10c by 
\citep{Fressin11}, as well as the planets in the Kepler 62 system \citep{Borucki13}. 
False alarm probabilities are inferred to be dramatically lower for cases where 
multiple candidate planets transit the same star. Among the gray points in 
Figure \ref{fig:kepler-multis} there is very likely only a relatively small admixture of 
false alarms.

\subsection{Results and Implications}

Aside from the sheer increase in the number of transiting planets that are known, 
the string of transit discoveries over the past six years have been of fundamentally 
novel importance. In particular, transit detections have enabled the study of both 
planets and planetary system architectures for which there are no solar system 
analogs. A brief tally of significant events logged in order of discovery year 
might include ({\bf i}) Gliese 436~b \citep{Gillon07} the first transiting 
Neptune-sized planet and the first planet to transit a low-mass star, 
({\bf ii}) HD 17156~b the first transiting planet with a large orbital 
eccentricity (e=0.69) and an orbital period ($P=21d$) that is substantially larger 
than the $2\,{\rm d}<P<5\,{\rm d}$ range occupied by a typical hot Jupiter 
\citep{Barbieri07},  ({\bf iii}) CoRoT 7~b \citep{Leger09} and Gliese 1214b 
\citep{Charbonneau2009} the first transiting planets with masses in the 
so-called ``super-Earth'' regime $1\,{\rm M}_{\oplus}<M<10\,{\rm M}_{\oplus}$, 
({\bf iv}) Kepler 9b and 9c \citep{Holman10} the first planetary system to show 
tangible transit timing variations, as well as the first case of transiting 
planets executing a low-order mean motion resonance, ({\bf v}) Kepler 22b, the 
first transiting planet with a size and an orbital period that could potentially 
harbor an Earth-like environment \citep{Borucki12}, and ({\bf vi}) the Kepler 62 
system \citep{Borucki13}, which hosts at least five transiting planets orbiting 
a K2V primary. The outer two members, Planet ``e'' with $P=122~{\rm d}$ and 
Planet ``f'' with $P=267~{\rm d}$, both have $1.25R_{\oplus}<R_{\rm p}<2R_{\oplus}$, 
and receive $S=1.2\pm0.2\,S_{\odot}$ and  $S=0.4\pm0.05\,S_{\odot}$ of Earth's solar 
flux respectively.

\begin{figure}[h]
\includegraphics[width=\columnwidth]{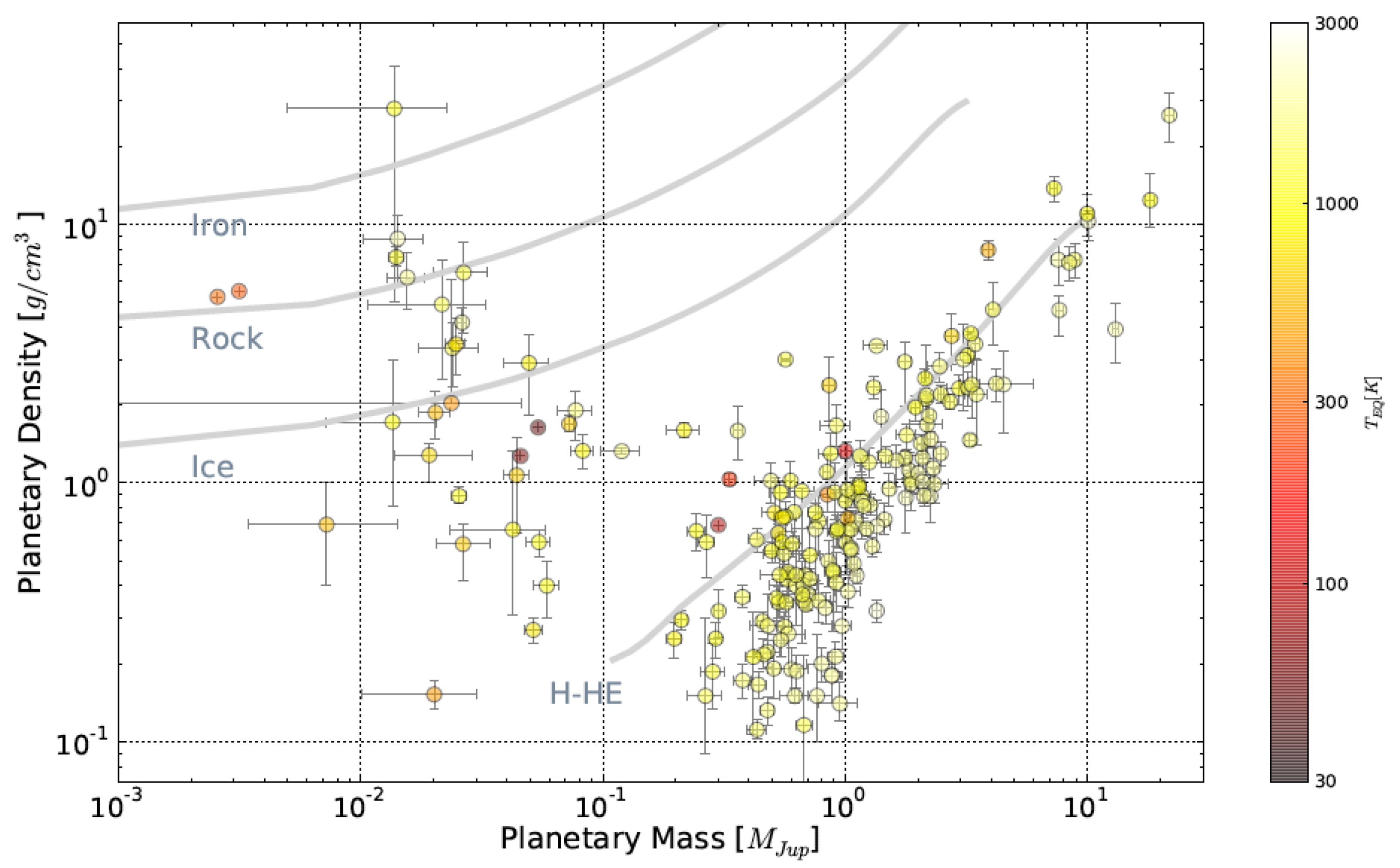}
\caption{Density-Mass diagram for planets with well-determined masses and radii. 
Planets are color-coded by the equilibrium temperature, 
$T_{\rm eq}=({R_{\star}^{1/2} T_{\star} )/({(2a)^{1/2}(1-e^{2})^{1/8}}})$, that 
they would have if they were zero-albedo black-bodies re-radiating from the full 
planetary surface area. The solar system planets more massive than Mars are 
included in the plotted aggregate. Gray lines show expected $\rho(M_P)$ for 
planetary models of pure hydrogen-helium, pure water, pure silicate, and pure iron 
compositions. Planetary data are from www.exoplanets.org, accessed 08/15/2013.
\label{fig:planet_densities}}
\end{figure}

Bulk densities are measured for transiting planets with parent stars that are bright 
enough and chromospherically quiet enough to support Doppler measurement of 
$M_P\sin(i)$, and can also be obtained by modeling transit timing variations 
\citep{Fabrycky2012, Lithwick12}. Over 100 planetary densities (mostly for hot 
Jupiters) have been securely measured. These are plotted in Figure 
\ref{fig:planet_densities}, which hints at the broad outlines of an overall 
distribution. Figure \ref{fig:planet_densities} is anticipated to undergo rapid 
improvement over the next several years as more Kepler candidates receive mass 
determinations. It appears likely, however, that there exists a very broad range 
of planetary radii at every mass. For example, to within errors, planets with 
$M_P \sim6M_{\oplus}$ appear to range in radius by a factor of at least three. 
While a substantial number of short-period giant planets are inflated by unknown 
energy source(s) \citep{Batygin10}, compositional variations are at least capable 
of explaining the observed range of radii for planets with $M_P<0.2M_{\rm Jup}$ 
\citep{Fortney2007}. The mass-density distribution (and by extension, the composition 
distribution) of extrasolar planets as a function of stellocentric distance is an 
important outcome of the planet formation process. It is still entirely unclear 
whether planets with $P<100\,{\rm d}$ that have no solar system analogues are the 
product of migration processes \citep{LinIda04} or of {\it in-situ} formation 
\citep{Chiang13}. More high quality measurements of transiting planets will be 
required to resolve the puzzle.

The large number of candidate multiple transiting planet systems indicate that 
co-planar architectures are the rule for planets with $P<100\,{\rm d}$ in the size 
range of $R_{\rm p} \sim$ 1.5 -- 6 R$_\oplus$ \citep{Moorhead11}. The inclination 
dispersion of most candidate systems with two or more transiting planets appears 
to have a median between 1--3$^{\circ}$. Candidate planets in multiple-transit 
systems, furthermore, are invariably in dynamically stable configurations 
when imbued with reasonable 
mass-radius relations \citep{Lissauer2011}. Nature has therefore produced a galactic 
planetary census that is extraordinarily well-suited to detection and characterization 
via the transit method. The advent of the new space missions, in concert with JWST's 
potential for atmospheric characterization of low-mass planets \citep{Deming09} 
indicate that transits will remain at the forefront for decades to come.

Finally, transit detection is unique in that it democratizes access to cutting-edge 
research in exoplanetary science. Nearly all of the highly-cited ground-based 
discoveries have been made with small telescopes of aperture $d<1$m. Amateur observers 
were co-discoverers of the important transits by HD 17156b \citep{Barbieri07}, and 
HD 80606b \citep{Garcia09}, and citizen scientists have discovered several planets 
to date in the Kepler data under the auspices of the Planet Hunters 
project \citep{Fischer12, Lintott13, Schwamb13}

\section{Direct Imaging Techniques} 
The field of exoplanets is almost unique in astronomical science in that the subjects are almost all studied indirectly, through their effects on more visible objects, rather than being imaged themselves. The study of the dominant constituents of the universe (dark energy and dark matter) through their gravitational effects is of course another example. Direct imaging of the spatially resolved planet is a powerful complement to the other techniques described in this chapter. It is primarily sensitive to planets in wide orbits $a > 5 $\,AU, and since photons from the planets are recorded directly, the planets are amenable to spectroscopic or photometric characterization. However, direct detection also represents a staggering technical challenge. If a twin to our solar system were located at a distance of 10 pc from the Earth, the brightest planet would have only $\sim 10^{-9}$ the flux of the parent star, at an angular separation of 0.5 arcseconds. 

In spite of this challenge, the field has produced a small number of spectacular successes: the images and spectra of massive  ($ >1000$ \mearth )  young self-luminous planets. The advent of the first dedicated exoplanet imaging systems should lead to rapid progress and surveys with statistical power comparable to ground-based Doppler or transit programs. In the next decade, space-based coronagraphs will bring mature planetary systems into reach, and some day, a dedicated exoplanet telescope may produce an image of an Earth analog orbiting a nearby star.  

\subsection{Limitations to high-contrast imaging}

The greatest challenge in direct imaging is separating the light of the planet from residual scattered light from the parent star. This can be done both optically -- removing the starlight before it reaches the science detector -- and in post-processing, using feature that distinguishes starlight from planetary light. 

\subsubsection{High-contrast point spread function, coronagraphs, and adaptive optics}
Even in the absence of aberrations, the images created by a telescope will contain features that will swamp any conceivable planet signal. The point spread function (PSF), as the name implies, is the response of the telescope to an unresolved point source. In the case of an unaberrated telescope, the PSF is the magnitude squared of the Fourier transform of the telescope aperture function. For an unobscured circular aperture, the diffraction pattern is the distinctive Airy rings. (The one-dimensional equivalent would represent the telescope as a top hat function, whose Fourier transform is a sinc, giving a central peak and oscillating sidelobes.) More complex apertures will have more complex diffraction patterns. 

Removing this diffraction pattern is the task of a coronagraph. Originally developed by \citet{L1939} to allow small telescopes to study the coronae of the sun, chronographs employ optical trickery to remove the light from an on-axis star while allowing some of the flux from the off-axis planet to remain. A wide variety of approches have been developed \citep{k_Kuchner_Collins_Ridgway_2006}, far too many to enumerate here, though they can be divided into a broad families. The classical Lyot coronagraph blocks the on-axis source with a focal plane mask, followed by a pupil-plane Lyot mask that blocks the light diffracted by the focal plane \citep{L1939, Makidon_Berkefeld_Kuchner_2001}. Apodizers operate by modifying the transmission of the telescope so that the Fourier transform has substantially less power in the sidelobes; a nonphysical example would be a telescope whose transmission was a smoothly-varying gaussian, which would result in a purely gaussian PSF. In more practical designs, apodization is implemented through binary ''shaped pupil" masks \citep{Vanderbei_Spergel_Littman_2003} and sharply reduce diffraction over a target region at a significant cost in throughput.  Hybrid Lyot approaches use pupil-plane apodization \citep{S11} or complicated focal-plane masks \citep{Kuchner_Traub_2002} to boost the performance of the classic Lyot. Phase-induced amplitude apodization uses complex mirrors to create the tapered beam needed to suppress diffraction without a loss in throughput \citep{rtinache_Ridgway_Woodruff_2005}. A particularly promising new technique creates an optical vortex in the focal plane \citep{an_Tabiryan_Mawet_Serabyn_2013} removing the diffracted light almost perfectly for an on-axis source in a unobscured aperture. Many more complex coronagraphs exist - see \cite{guyon2006} for discussion. Typically, the best coronagraphs remove diffraction down to the level of $10^{-10}$ at separations greater than the inner working angle (IWA), typically $2-4 \lambda / D$.

Light is also scattered by optical imperfections - wavefront errors induced by the telescope, camera, or atmospheric turbulence. Even with a perfect coronagraph, atmospheric turbulence, which typically is many waves of phase aberration produces a PSF that completely overwhelms any planetary signal. Even in the absence of atmospheric turbulence, small wavefront errors from e.g., polishing marks will still scatter starlight. These can be partially corrected through adaptive optics - using a deformable mirror (DM), controlled by some estimate of the wavefront, to correct the phase of the incoming light. In the case of small phase errors, a Fourier relationship similar to that for diffraction exists between the wavefront and PSF - see \citet{perrin2003} and \citet{guyon2006} for discussion and examples. A useful figure of merit for adaptive optics correction is the Strehl ratio, defined as the ratio of the peak intensity of the measured PSF to the theoretical PSF for an equivalent unaberrated telescope. With current-generation adaptive optics systems, Strehl ratios of 0.4-0.8 are common in K band - meaning that 60-80 percent of the scattered light remains uncorrected. 

The halo of light scattered by wavefront errors is particularly troublesome because it does not form a smooth background, but is broken up into a pattern of speckles. In monochromatic light these speckles resemble the diffraction-limited PSF of the telescope, and hence are easily confused with the signal from a planet. As a result, high-contrast images are usually nowhere near the Poisson limit of photon noise but instead limited by these speckles. Uncorrected atmospheric turbulence produces a halo of speckles that rapidly evolve; static or quasi-static wavefront errors, such as adaptive optics miscalibrations, produce slowly evolving speckles that mask planetary signals.

\subsubsection{Post-processing}

These speckle patterns can be partially mitigated in post-processing. Such PSF subtraction requires two components. First, there must be some distinction between a planetary signal and the speckle pattern - some {\it diversity}. Examples include wavelength diversity, where the wavelength dependence of the speckle pattern differs from that of the planet; rotational diversity, in which the telecope (and associated speckle pattern) rotates with respect to the planet / star combination \citep{Marois06}; or observations of a completely different target star. Such reference PSFs will never be a perfect match, as the PSF evolves with time, temperature, star brightness, and wavelength. The second component needed for effective PSF subtraction is an algorithm that can construct the ``best" PSF out of a range of possibilities.  With a suitable library of PSFs, least-squares fitting \citep{lafeniere2007} or principal components analysis can assemble synthetic PSFs and enhance sensitivity to planets by a factor of 10-100. 

\subsection{Imaging of self-luminous planets}
\begin{figure}[!htb]
\begin{center}
\includegraphics[width=0.98\columnwidth]{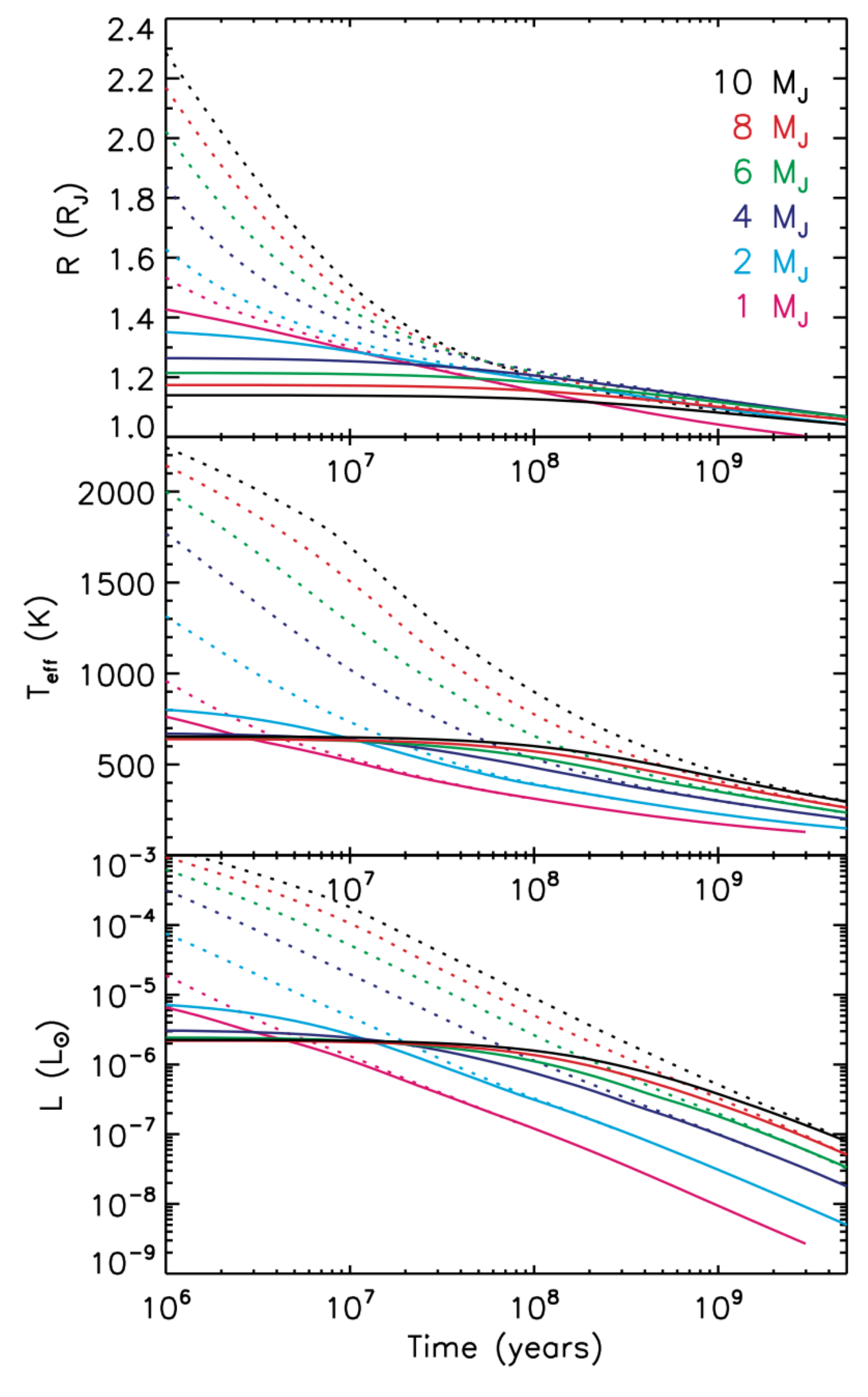}
\caption{\label{fig:Marley_fig4} Reproduction of Fig 4 from \citet{M07} showing the model radius, temperature and luminosity of young Jupiters as a function of time since the beginning of their formation. Different colors reflect different planetary masses. Dotted lines indicate ``hot start'' planets, where adiabatic formation retains most of the initial energy and entropy; solid lines indicate ``cold start'', where accretion through a shock (as in the standard core accretion paradigm) results in loss of entropy. In either case, planets are singnificantly easier to detect at young ages. }
\end{center}
\end{figure}

With these techniques applied to current-generation systems, planets with brightness $\sim 10^{-5}$ can be seen at angular separations of $\sim 1.0$ arcseconds. This is far from the level of sensitivity needed to see mature Jupiter-like planets. Fortunately, planets are available that are much easier targets. When a planet forms, signficant gravitational potential energy is available. Depending on the details of initial conditions, a newly-formed giant planet may have an effective temperature of 1000-2000 K \citep{M07} and a luminosity of $10^{-5}$ to $10^{-6}\, L_\odot$ (Fig. \ref{fig:Marley_fig4}). As with the brown dwarfs, a large fraction of this energy could be released in the near-infrared, bringing the planet into the detectable range. Such planets remain detectable for tens of millions of years. Several surveys have targeted young stars in the solar neighborhood for exoplanet detection (\cite{liu2010}, \cite{gdps2007}, \cite{chauvin2010}, benefitting from the identification of nearby young associations composed of stars with ages 8-50 Myr \citep{ZS04}.  Most of these surveys have produced only non-detections, with upper limits on the number of giant planets as a function of semi-major axis that exclude large numbers of very-wide orbit (50 AU) planets.

\begin{figure}[!htb]
\begin{center}
\includegraphics[width=0.95\columnwidth]{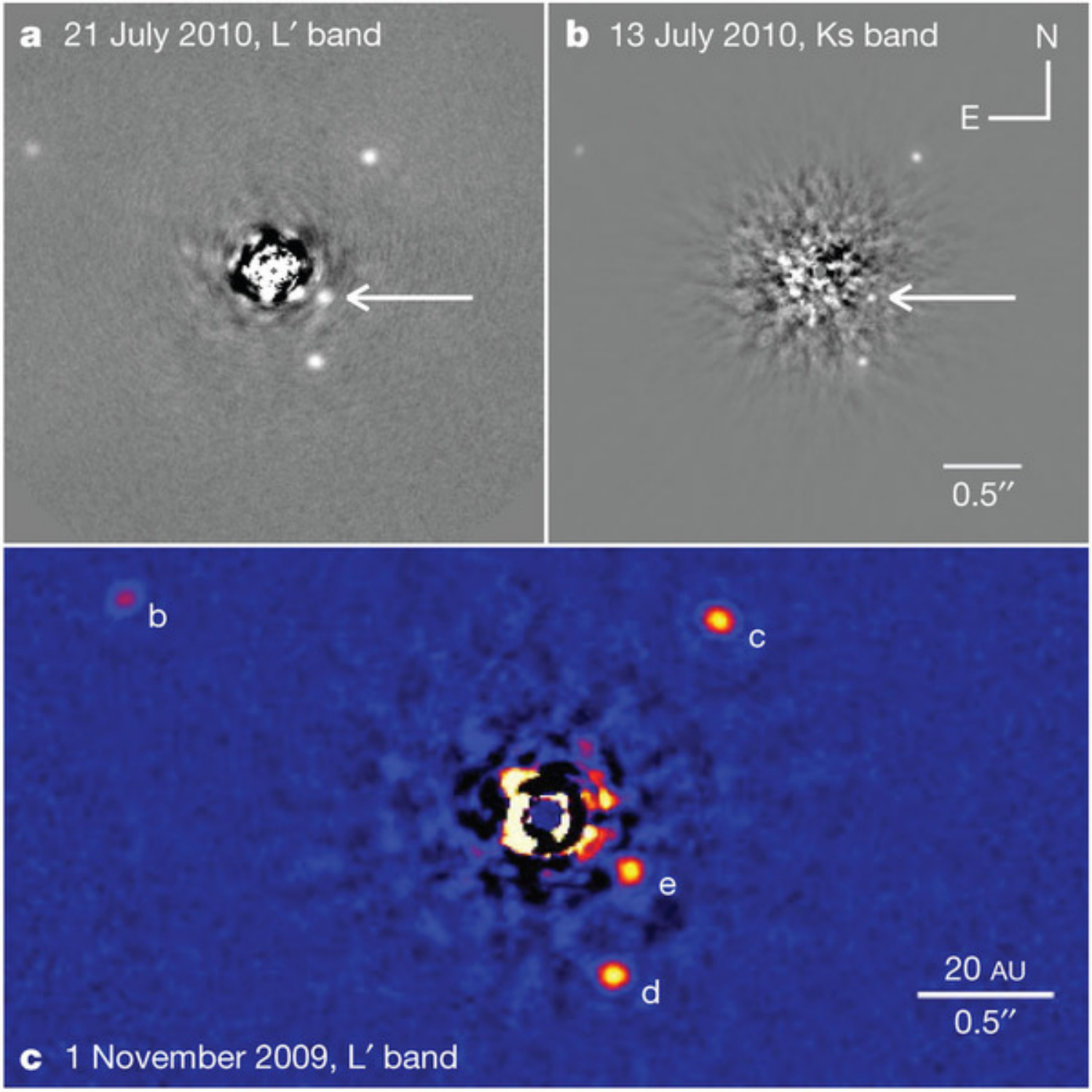}
\caption{\label{fig:nature09684}Near-infrared Keck adaptive optics images of the HR8799 system from \cite{marois2010}. Four giant planets, 3 to 7 times the mass of Jupiter, are visible in near-infrared emission.The residual speckle pattern after PSF subtraction can be seen in the center of each image.}
\end{center}
\end{figure}

A handful of spectacular successes have been obtained. One of the first detections was a 5 Jupiter-mass object that was orbiting not a star but a young brown dwarf, 2M1207B \citep{llet_Song_Beuzit_Lowrance_2004}. A spectacular example of planetary companions to a main-sequence star is the HR8799 multi-planet system (Figure \ref{fig:nature09684}). This consists of four objects near a young F0V star, orbiting in counterclockwise directions. The object's luminosities are well-constrained by broad-band photometry \citep{marois2008,currie2011}. Estimates of the planetary mass depend on knowledge of the stellar age - thought to be 30 Myr \citep{marois2010, baines2012} and initial conditions; for 'hot start' planets the masses are 3-7 times that of Jupiter. Multi-planet gravitational interactions provide a further constraint on the mass \citep{marois2010, fabrycky2010}, excluding massive brown dwarf companions. Other notable examples of directly imaged exoplanets include the very young object 1RXS J1609b \citep{lafreniere2010}, the cool planet candidate GJ504B \citep{gj504}, and the planet responsible for clearing the gap inside the Beta pictoris disk \citep{lagrange2010}. A candidate optical HST image of an exoplanet was reported orbiting Fomalhaut \citep{kalas2008}, but very blue colors and a belt-crossing orbit \citep{kalas2013} indicate that what is seen is likely light scattered by a debris cloud or disk (that may still be associated with a planet).

The photometric detections of self-luminous planets have highlighted the complexities of modeling the atmospheres of these objects. Although they are similar to brown dwarfs, many of the directly imaged planets have temperatures that place them in the transitional region between cloud-dominated L dwarfs and methane-dominated T dwarfs - a change that is poorly understood even for the well-studied brown dwarfs. Cloud parameters in particular can make an enormous difference in estimates of properties like effective temperature and radius (see supplementary material in \cite{marois2008} and subsequent discussion in \cite{barman2011}, \cite{marley2012}, \cite{currie2011}, and discussion in the chapter by Madhusudhan et al. in this volume.  

\begin{figure}[!htb]
\begin{center}
\includegraphics[width=\columnwidth]{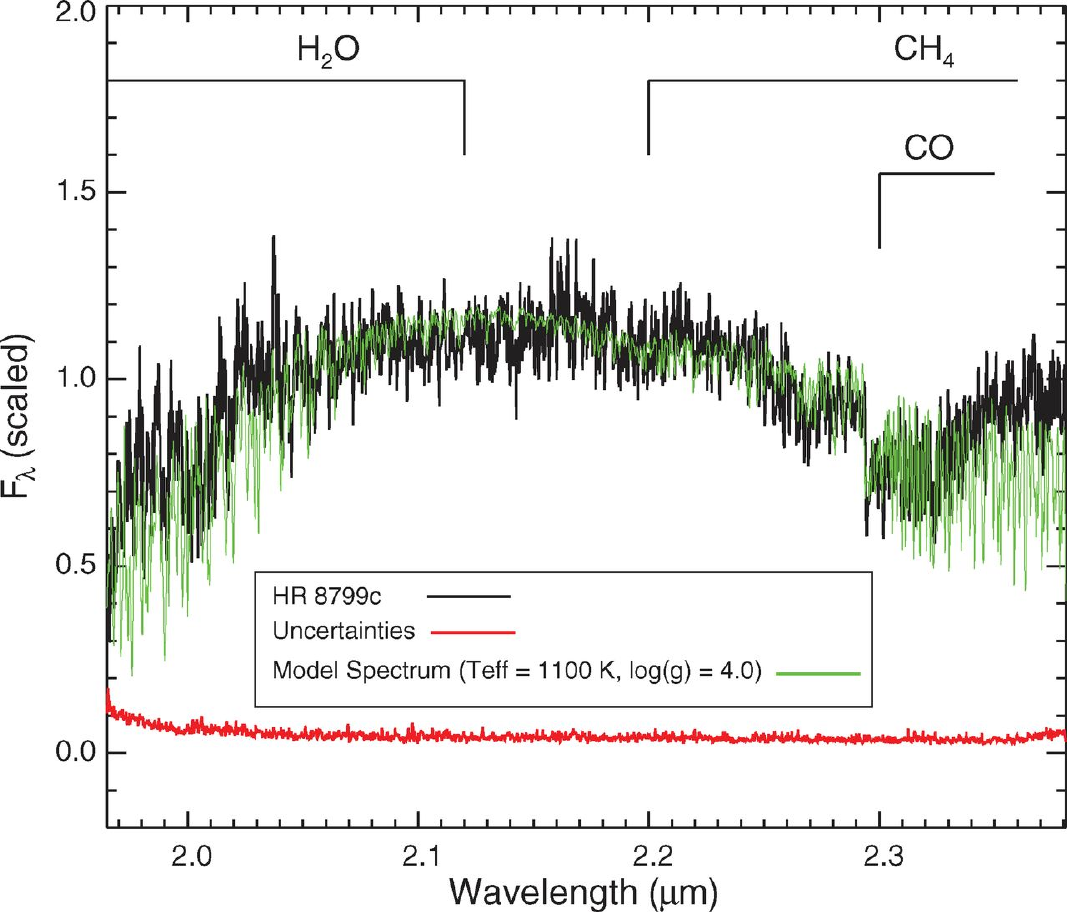}
\caption{\label{fig:F2large}High-resolution spectrum of the extrasolar planet HR8799c taken with the OSIRIS spectrograph and the Keck adaptive optics system, reproduced from \cite{y_Barman_Macintosh_Marois_2013}. Residual speckle noise changes the overall spectral shape (e.g., the upturn at the long wavelength end) but does not inject narrow features - the CO break is clearly detected as are many individual CO and H2O lines, while methane is absent. }
\end{center}
\end{figure}

If a planet can be clearly resolved from its parent star, it is accessible not only through imaging but also spectroscopically. Integral field spectrographs are particularly well suited to this, e.g., \citet{Dekany_Fergus_Hale_et_al__2013, y_Barman_Macintosh_Marois_2013}, since they also capture the spectrum of neighboring speckle artifacts, which can be used to estimate the speckle contamination of the planet itself. Spectra show that the self-luminous planets do (as expected) have low gravity and distinct atmospheric structure from brown dwarfs. In some cases, spectra have sufficiently high SNR that individual absorption features (e.g., of CO) can be clearly resolved \citep{y_Barman_Macintosh_Marois_2013}, allowing direct measurements of atmospheric chemistry and abundances (Figure \ref{fig:F2large}).

\subsection{Future ground and space-based facilities}     Most direct imaging of exoplanets to date has taken place with traditional instruments attached to general-purpose AO systems, such as the NIRC2 camera on the Keck II telescope or NACO on the VLT. In fact, for most of these observations, the presence or absence of a coronagraph has had little effect on sensitivity, which is dominated by wavefront errors uncorrected by the AO system. Some sensitivity enhancement has come from dedicated exoplanet imaging cameras, employing techniques like dual-channel imaging, in combination with conventional adaptive optics \citep{kemer_Chun_Ftaclas_et_al__2013, rrie_McElwain_Itoh_et_al__2013}. The combination of pyramid wavefront sensing and adaptive secondary mirrors on the LBT and Magellan telescopes has shown excellent high-contrast performance \citep{skemer2012}. 

However, to significantly increase the number of imaged exoplanets will require dedicated instruments that combine very high-performance adaptive optics, suitable coronagraphs, and exoplanet-optimized science instruments such as low spectral resolution diffraction-limited integral field spectrographs.     The first such instrument to become operational is the Project 1640 coronagraphic IFS \citep{Dekany_Fergus_Hale_et_al__2013}, integrated with a 3000-actuator AO system on the 5-m Hale telescope. The Subaru Coronagraphic Extreme AO System \citep[SCExAO;][]{yon_Clergeon_Garrel_Blain_2012} is a 2000-actuator AO system that serves as a testbed for a wide variety of advanced technologies including focal-plane wavefront sensing and pupil-remapping coronagraphs. Finally, two facility-class planet imagers will be operational in 2014 on 8-m class telescopes - the Gemini Planet Imager \citep{Dillon_Doyon_Dunn_et_al__2012} and the VLT SPHERE facility \citep{Beuzit2008, uey_Charton_Dohlen_et_al__2012}. Both have 1500 actuator AO systems, apodized-pupil Lyot coronagraphs, and integral field spectrographs. (SPHERE also incorporates a dual-channel IR imager and a high-precision optical polarimeter.) Laboratory testing and simulations predict that they will achieve on-sky contrasts of better than $10^6$ at angles of 0.2 arcseconds, though with the limitation of requiring bright stars ($I<8$ mag for GPI, $V<12$ mag for SPHERE ) to reach full performance. Both instruments will be located in the southern hemisphere, where the majority of young nearby stars are located. Simulated surveys \citep{arois_Poyneer_Wiktorowicz_2011} predict that GPI could discover 20-50 Jovian planets in a 900-hour survey.          

Direct detection instruments have also been proposed for the upcoming 20-40m Extremely Large Telescopes. These instruments exploit the large diameters of the telescope to achieve extremely small inner working angles (0.03 arcseconds or less), opening up detection of protoplanets in nearby star forming regions orbiting at the snow line \citep{macintosh2006}, or reflected light from mature giant planets close to their parent star \citep{kasper2010}. At their theoretical performance limits, such telescopes could reach the contrast levels needed to detect rocky planets in the habitable zones of nearby M stars, though reaching that level may present insurmountable technical challenges. \citep{guyon2012}. 

A coronagraphic capability has  been proposed for the 2.4m AFTA WFIRST mission \citep{Spergel13}. Due to the obscured aperture and relative thermal stability of the telescope, it would likely be limited to contrasts of $10^{-9}$ at separations of 0.1 or 0.2 arcseconds, but this would still enable a large amount of giant-planet and disk science, including spectral charcterization of mature giant planets. 

Direct detection of an Earth-analog planet orbiting a solar-type star, however, will almost certainly require a dedicated space telescope using either an advanced coronagraph - still equiped with adaptive optics - or a formation-flying starshade occulter.

\section{Microlensing}

\subsection{Planetary Microlensing}

\subsubsection{Microlensing Basics}

\begin{figure}[!htb]
\includegraphics[width=0.98\columnwidth]{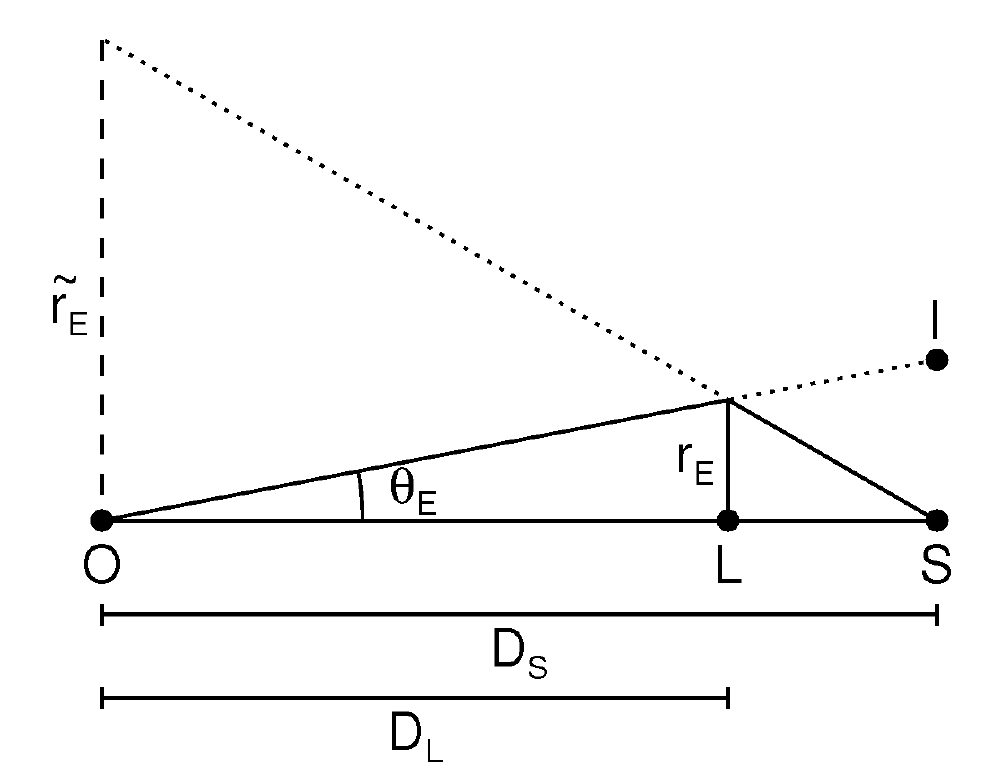}
\caption{Basic geometry of microlensing.\label{fig:geom}}
\end{figure}

A microlensing event occurs when two stars at different distances pass
within $\sim 1$ mas of each other on the plane of the sky
\citep{Gaudi12}. Light from the source star `S' is bent by the lens
star `L', so that the observer `O' sees the the image `I' instead of
the true source (see Fig. \ref{fig:geom}). If the source and the
lens are perfectly aligned along the line of sight, the source is
lensed into a ring \citep{Chwolson24,Einstein36,Renn97}, called an
Einstein ring whose angular size is given by:
\begin{equation}
\label{eqn:thetaE}
\theta_{\rm E} = \sqrt{\kappa M_{\rm L} \pi_{\rm rel}} \sim 0.3\, \mathrm{ mas }
\end{equation}
for typical values of the lens mass ($M_{\rm L} = 0.5 M_{\odot}$),
lens distance ($D_{\rm L} = 6\,$ kpc), and source distance ($D_{\rm S}
= 8\,$ kpc). In Equation \ref{eqn:thetaE}, $\pi_{\rm rel} = (1 {\rm
  AU}/D_{\rm L})-(1 {\rm AU}/D_{\rm S})$ is the trigonometric parallax
between the source and the lens, and $\kappa = 8.14\, $mas
M$_{\odot}^{-1}$.

\begin{figure}[!htb]
\includegraphics[width=\columnwidth]{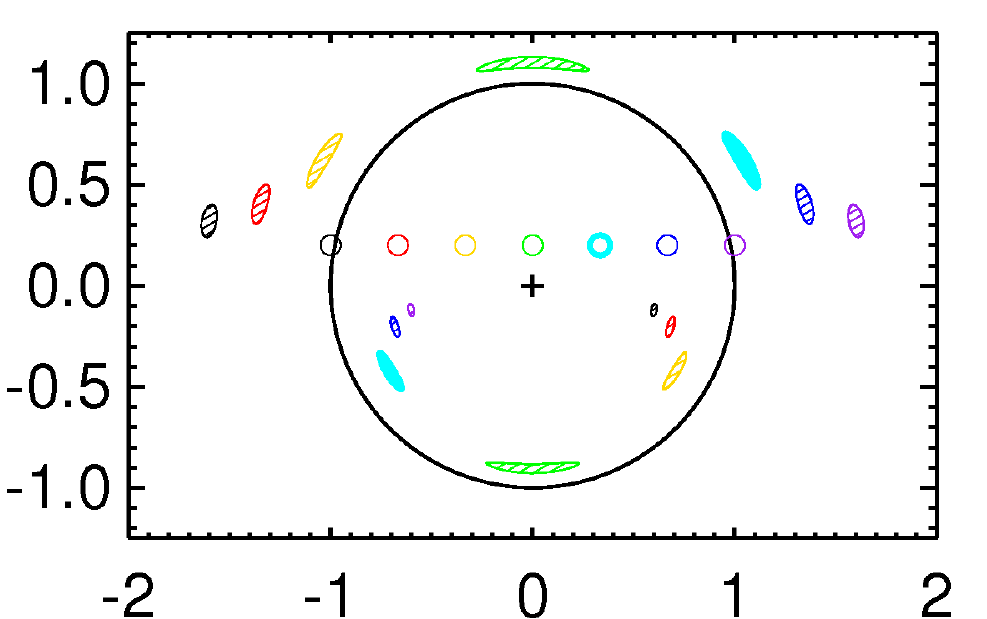}
\caption{Images of a lensed source star. The position of the source is
  indicated by the small circles. The filled ovoids show the lensed
  images for each source position. The large black circle shows the
  Einstein ring. The lens star is at the origin, marked by the
  plus.\label{fig:images}}
\end{figure}

If the source is offset from the lens by some small amount, it is
lensed into two images that appear in line with the source and the
lens, and close to the Einstein ring as in Figure
\ref{fig:images}. Because the size of the Einstein ring is so small,
the two images of the source are unresolved and the primary observable
is their combined magnification
\begin{equation}
\label{eqn:mag}
A=\frac{u^2+2}{u\sqrt{u^2+4}},
\end{equation}
where $u$ is the projected separation between the source and the lens
as a fraction of the Einstein ring. Since the source and the lens are
both moving, $u$ (and so $A$) is a function of time.

\subsubsection{Types of Planetary Perturbations}

If planets are gravitationally bound to a lensing star, the planet can
be detected if one of the source images passes over or near the
position of the planet. This creates a perturbation to the
microlensing light curve of the host star.  Because the images
generally appear close to the Einstein ring, microlensing is most
sensitive to planets with projected separations equal to the physical
size of the Einstein ring in the lens plane, $r_{\rm E} =
\theta_{\rm E}D_{\rm L}$.

Another way to think about this is to consider the magnification
map. The magnification of the source by a point lens can be calculated
for any position in space using Equation \ref{eqn:mag}, giving a
radially symmetric magnification map. The source then traces a path
across this map creating a microlensing event whose magnification
changes as a function of time (and position).  The presence of the
planet distorts the magnification map of the lens and causes two or
more caustics to appear as shown by the red curves in Figure
\ref{fig:lcs}a. A perfect point source positioned at a point along the
caustic curve will be infinitely magnified. In order to detect the
planet, the source trajectory must pass over or near a caustic caused
by the planet \citep{MaoPaczynski91,GouldLoeb92,GriestSafizadeh98}.

There are two kinds of perturbations corresponding to the two sets of
caustics produced by the planet. The ``planetary caustic'' is the
larger caustic (or set of caustics) unassociated with the position of
the lens star (right side of Fig. \ref{fig:lcs}a). The ``central
caustic'' is much smaller than the planetary caustic and is located at
the position of the lens star (left side of Fig.
\ref{fig:lcs}a). Figure \ref{fig:lcs} shows two example source
trajectories, their corresponding light curves, and details of the
planetary perturbation in a planetary caustic crossing. As the mass
ratio, $q$, decreases, so does the duration of the planetary
perturbation. In addition, the detailed shape of the perturbation
depends on the size of the source star relative to the size of the
Einstein ring, $\rho$.

\subsubsection{Planet Masses from Higher-Order Effects}

The fundamental observable properties of the planet are the mass ratio
between the planet and the lens star, $q$, and the projected
separation between the planet and the lens star as a fraction of the
Einstein ring, $s$. Hence, while $q\leq10^{-3}$ definitively
identifies the companion to the lens as a planet, its physical
properties cannot be recovered without an estimate of $M_{\rm L}$ and
$D_{\rm L}$. However, if $\theta_{\rm E}$ and $\tilde{r}_{\rm E}$ (the
size of the Einstein ring in the observer plane) can be measured, it
is possible to obtain measurements of $M_{\rm L}$ and $D_{\rm L}$ (see
Fig. \ref{fig:geom}) and hence, the physical mass and projected
separation of the planet: $m_{\rm p} = qM_{\rm L}$ and $a_{\perp} =
s\theta_{\rm E}D_{\rm L}$.  These variables can be measured from
higher-order effects in the microlensing light curve. If finite-source
effects are observed (c.f. Fig. \ref{fig:lcs}e), $\theta_{\rm E}$ is
measured since $\rho=\theta_{\star}/\theta_{\rm E}$ and the angular
size of the source, $\theta_{\star}$, can be determined from the
color-magnitude diagram \citep{Yoo04b}. Finally, as the Earth orbits
the Sun, the line of sight toward the event changes giving rise to
microlens parallax \citep{Gould92a,GouldMB94}, allowing a measurement
of $\tilde{r}_{\rm E}$:
\begin{equation}
\pi_{\rm E} = \frac{1 {\rm AU}}{\tilde{r}_{\rm E}}=\frac{\pi_{\rm
    rel}}{\theta_{\rm E}}.
\end{equation}

\subsubsection{Microlensing Degeneracies and False-Positives}

In microlensing the most common degeneracy is that planets with
separation $s$ produce nearly identical central caustics as planets
with separation $s^{-1}$ \citep{GriestSafizadeh98}. For planetary
caustics, this is not a major problem since $s$ (where $s$ is larger
than the Einstein ring) produces a ``diamond''-shaped caustic whereas
$s^{-1}$ produces a pair of ``triangular'' caustics
\citep{GaudiGould97}. Additional degeneracies arise when higher-order
effects such as parallax and the orbital motion of the lens are
significant. In such cases, the exact orientation of the event on the
sky becomes important and can lead to both discrete and continuous
degeneracies in the relevant parameters \citep[e.g. ][]{Gould04,Skowron11}.

\begin{figure}[!htb]
\includegraphics[width=0.95\columnwidth]{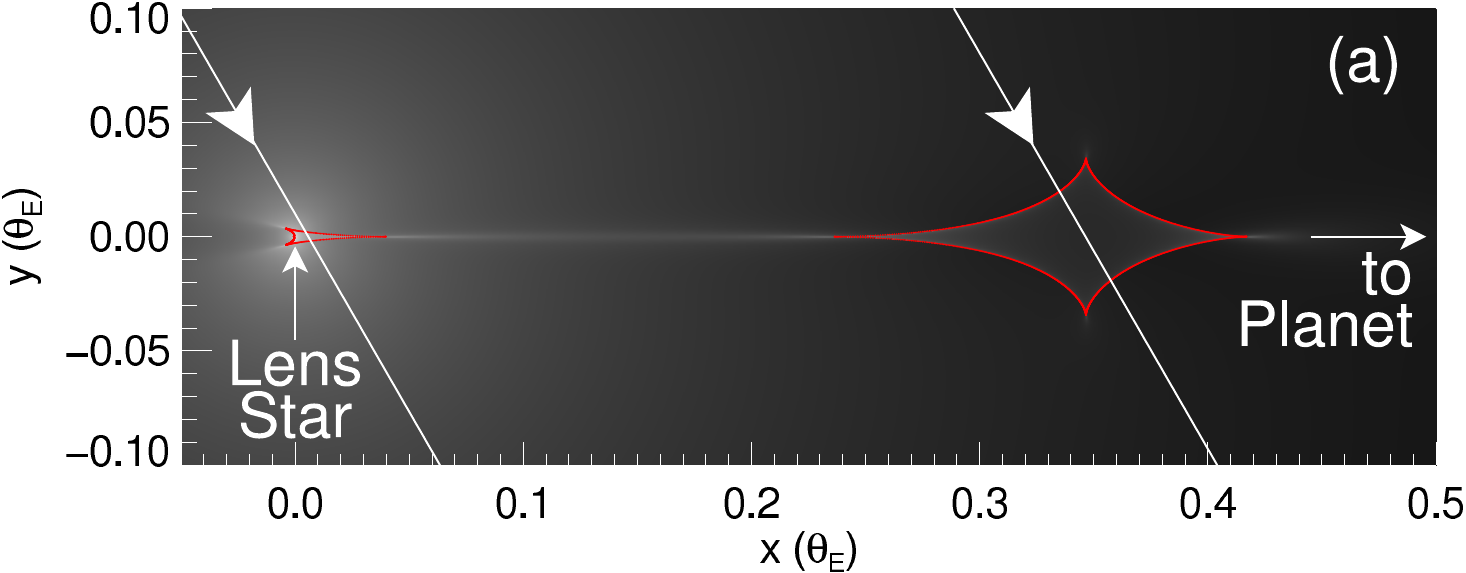}
\includegraphics[width=0.95\columnwidth]{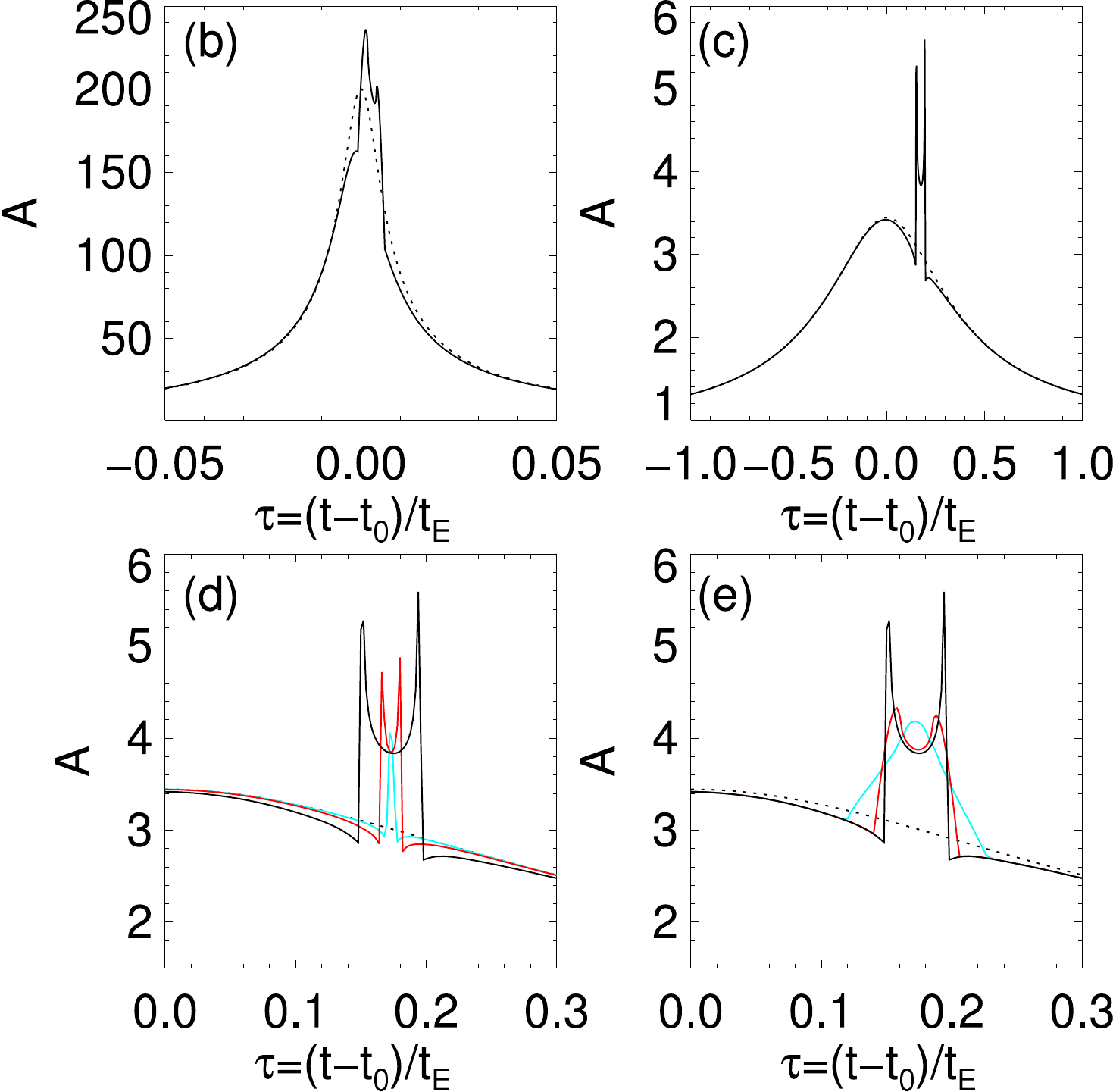}
\caption{
(a) Magnification map for a planet with $q=0.001$ and $s=1.188$ and a
source size $\rho=0.001$. The red lines indicate the caustics. Two
example source trajectories are shown. The scale is such that the
Einstein ring is a circle of radius 1.0 centered at (0,0). The planet
is located at (1.188,0), just outside the Einstein ring (off
the right-hand side of the plot).
(b) Light curve corresponding to the left-hand source trajectory (a
central caustic crossing). The dotted line shows the corresponding
light curve for a point lens.
(c) Light curve corresponding to the right-hand source trajectory (a
planetary caustic crossing).
(d) Detail of (c) showing the variation in the planetary signal for
different values of $q=10^{-3}, 10^{-4}, 10^{-5}$ (black, red, cyan).
(e) The variation in the planetary signal for different values of
$\rho=0.001, 0.01, 0.03$ (black, red, cyan).
\label{fig:lcs}}
\end{figure}

False positives are rare in microlensing events in which the source
crosses a caustic. Because the magnification diverges at a caustic,
this produces a discontinuity in the slope of the light curve, which
is very distinctive (see Fig. \ref{fig:lcs}). However, in events without
caustic crossings, planetary signals can be mimicked by a binary
source \citep{Gaudi98,Hwang13}, orbital motion of the lens
\citep[e.g. ][]{Albrow00b}, or even starspots
\citep[e.g. ][]{Gould13}. Often multi-band data can help distinguish
these scenarios as in the case of starspots or lensing of two sources
of different colors.

\subsection{Microlensing Observations in Practice}

The first microlensing searches were undertaken in the late 1980s,
primarily as a means to find Massive Compact Halo Objects \citep[a
  dark matter candidate;][]{Alcock92,Aubourg93}. These searches were
quickly expanded to include fields toward the galactic bulge to search
for planets and measure the mass function of stars in the inner galaxy
\citep{Paczynski91,Griest91b}. One million stars must be observed to
find one microlensing event, so the first surveys focused on simply
detecting microlensing events. These surveys typically observed each
field between once and a few times per night.  However, the timescale
of the planet is much shorter: a day or two for a Jupiter down to an
hour for an Earth-mass planet. Hence, followup groups target the known
microlensing events to obtain the higher cadence observations necessary
to detect planets.

In practice, it is not possible to followup all microlensing events,
so the first priority is placed on the high-magnification events
($A\gtrsim50$), i.e., the central caustic crossing events. Not
only can the time of peak sensitivity to planets be predicted (around
the time of maximum magnification), but these events are much more
sensitive to planets than the average events, giving maximal
planet-yield for the available resources \citep{GriestSafizadeh98}.

To date, almost 20 microlensing planets have been published, most of
them found using the survey+followup method and in high magnification
events. Currently the main surveys for detecting microlensing events
are the Optical Gravitational Lens Experiment
\citep[OGLE;][]{Udalski03} and Microlensing Observations in
Astrophysics \citep[MOA;][]{Bond04} . Wise Observatory in Israel is
also conduction a microlensing survey toward the bulge
\citep{Gorbikov10,Shvartzvald12}. Combined these surveys now discover
over 2000 microlensing events each year. In addition, several
groups are devoted to following up these events. They are Microlensing
Follow-Up Network \citep[$\mu$FUN;][]{Gould06}, Microlensing Network
for the Detection of Small Terrestrial Exoplanets
\citep[MiNDSTEp;][]{Dominik10}, Probing Lensing Anomalies NETwork
\citep[PLANET;][]{Beaulieu06}, and RoboNet \citep{Tsapras09}. 

\subsection{Microlensing Planet Discoveries}

\subsubsection{Highlights}

\subsubsubsection{The First Microlensing Planet} The first
microlensing planet,
OGLE-2003-BLG-235/MOA-2003-BLG-53Lb, was a 2.6 $M_{\rm Jup}$ planet
discovered in 2003 by the OGLE and MOA surveys
\citep{Bond04}. Although it was discovered and characterized by
surveys, this planet was found in ``followup mode'' in which the MOA
survey changed its observing strategy to follow this event more
frequently once the planetary anomaly was detected.

\subsubsubsection{Massive Planets Around M-dwarfs} Many of the planets
discovered by microlensing have large mass ratios corresponding to
Jovian planets. At the same time, the microlensing host stars are
generally expected to be M dwarfs since those are the most common
stars in the galaxy. Specifically, there are two confirmed examples of
events for which the host star has been definitively identified to be
an M dwarf hosting a a super-Jupiter: OGLE-2005-BLG-071
\citep{Udalski05,Dong09_071} and MOA-2009-BLG-387
\citep{Batista11}. The existence of such planets is difficult to
explain since the core accretion theory of planet formation predicts
that massive, Jovian planets should be rare around M dwarfs
\citep{Laughlin04,Ida05}. However, it is possible they formed through
gravitational instability and migrated inward \citep{Boss06}.

\subsubsubsection{Multi-Planet Systems} Two of the microlensing events
that host planets, OGLE-2006-BLG-109 \citep{Gaudi08} and
OGLE-2012-BLG-0026 \citep{Han13}, have signals from two different
planets. The OGLE-2006-BLG-109L system is actually a scale model of
our solar system. The planets in this event are a Jupiter and a Saturn
analog, with both planets at comparable distances to those planets
around the Sun when the difference in the masses of the stars is taken
into account.

\subsubsubsection{Free-floating Planets} Because microlensing does not
require light to be detected from the lenses, it is uniquely sensitive
to detecting free-floating planets. Since $\theta_{\rm E}$ scales as
$M^{1/2}$, free-floating planets have extremely small Einstein rings
and hence give rise to short duration events ($\lesssim 1$ day). Based
on the analysis of several years of MOA survey data, \citet{Sumi11}
found that there are two free-floating Jupiters for every star.

\subsubsection{The Frequency of Planets Measured with Microlensing}

\begin{figure}[!htb]
\includegraphics[width=.95\columnwidth]{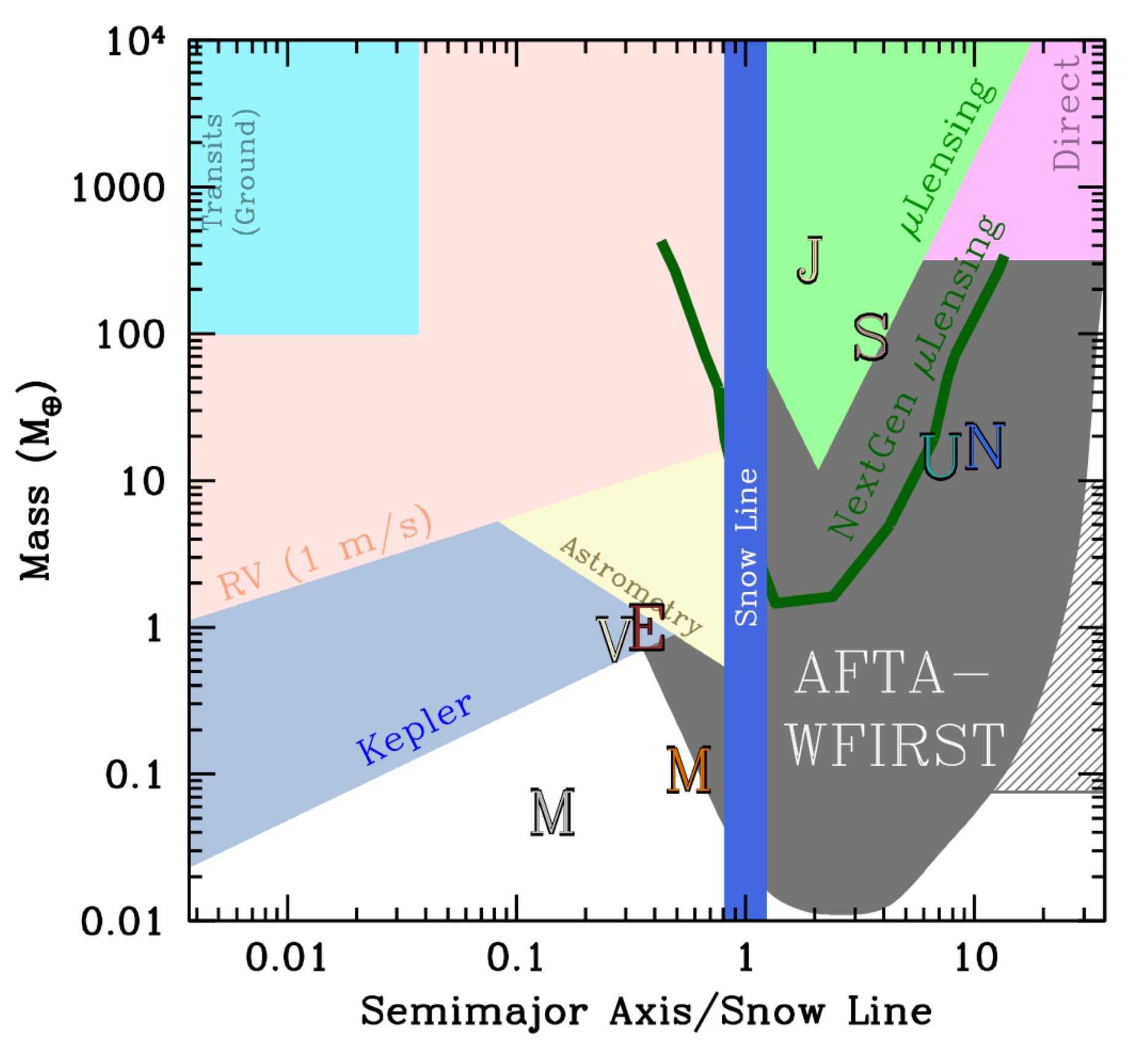}
\caption{Sensitivity of microlensing compared to other
  techniques. Figure courtesy B. Scott Gaudi and Matthew
  Penny.\label{fig:sens}}
\end{figure}

Figure \ref{fig:sens} compares the sensitivity of microlensing to
other techniques, where the semi-major axis has been scaled by the
snow line, $a_{\rm snow}=2.7 \mathrm{AU} (M_{\star}/M_{\odot}$).
The ``typical'' microlensing host is an M dwarf rather than a G dwarf,
so from the perspective of the core-accretion theory of planet
formation, the relevant scales are all smaller. In this theory, the
most important scale for giant planet formation is the location of the
snow line, which depends on stellar mass \citep{Ida04}. Microlensing
is most sensitive to planets at $1\, r_{\rm E}$, which is roughly 3
times $a_{\rm snow}$ for an M dwarf (i.e., $a_{\rm snow}\sim 1\,$ AU
and $r_{\rm E} \sim 3\,$ AU).

The frequency, or occurrence rate, of planets can be calculated by
comparing the sensitivities of individual events to the planets
detected. \citet{Gould10} analyzed high-magnification microlensing
events observed by $\mu$FUN from 2005-2008 and found $dN/(d\log q\,
d\log s)=0.31+/-0.15$ planets per dex$^2$ normalized at planets with
Saturn mass-ratios. \citet{Cassan12} also calculated the frequency of
planets using events observed by PLANET, including both high and low
magnification events. They found a similar planet frequency of $dN/(d
\log a\, d \log m_{\rm p}) = 10^{-0.62\pm0.22}(m_{\rm p}/M_{\rm
  Sat})^{0.73\pm0.17}$ normalized at Saturn-masses and flat as a
function of semi-major axis. Figures 8 and 9 in \citet{Gould10} compare
their result to the results from radial velocity for solar-type stars
\citep{Cumming08, Mayor09} and M dwarfs \citep{Johnson10}.

\subsection{The Future of Microlensing}

\subsubsection{Second-Generation Microlensing Surveys}

Advances in camera technology now make it possible to carry out the
ideal microlensing survey: one that is simultaneously able to monitor
millions of stars while also attaining a $\sim 15$ minute
cadence. Both OGLE and MOA have recently upgraded to larger
field-of-view cameras \citep[][]{Sato08,Soszynski12}. They have teamed
up with Wise Observatory in Israel to continuously monitor a few of
their fields \citep{Shvartzvald12}. In addition, the Korea
Microlensing Telescope Network \citep[KMTNet;][]{Park12} is currently
under construction. This network consists of three identical
telescopes in Chile, Australia, and South Africa, which will conduct a
high-cadence microlensing survey toward the galactic bulge. As these
second-generation surveys get established, they will dominate the
microlensing planet detections and the bulk of the detections will
shift to planetary caustic crossings. Although high-magnification
events are individually more sensitive to planets, they are very rare
compared to low-magnification events. Hence, the larger cross-section
of the planetary caustics will make low-magnification events the
dominant channel for detecting planets in the new surveys.

\subsubsection{Space-Based Microlensing}

The next frontier of microlensing is a space-based survey, which has
the advantages of improved photometric precision, the absence of
weather, and better resolution. The improved resolution that can be
achieved from space is a major advantage for characterizing the
planets found by microlensing. In ground-based searches the stellar
density in the bulge is so high that unrelated stars are often blended
into the $1^{\prime\prime}$ PSF. This blending makes it impossible to
accurately measure the flux from the lens star, and hence unless
higher-order microlensing effects are observed, it is difficult to
know anything about the lens. In space, it is possible to achieve a
much higher resolution that resolves this blending issue, allowing an
estimate of the lens mass based on its flux and hence, a measurement
of true planet masses rather than mass ratios.

The first microlensing survey satellite was proposed in
\citet{BennettRhie00, BennettRhie02}. Currently, a microlensing survey
for exoplanets has been proposed as a secondary science project for
the {\it Euclid} mission \citep{Penny13, BeaulieuTisserandBatista13}
and is a major component of the {\it WFIRST} mission
\citep{Spergel13}. The {\it WFIRST} mission is expected to detect
thousands of exoplanets beyond the snowline \citep{Spergel13}. The
parameter space probed by this mission is complementary to that probed
by the {\it Kepler} mission, which focused on detecting transits from
close-in planets (see Fig. \ref{fig:sens}).

\section{Astrometry}
\subsection{Introduction}
Steady advances in the 18th century improved the precision of stellar position 
measurements so that it was possible to measure the proper motions of stars, their 
parallax displacements due to Earth's motion around the Sun, and orbital motion 
caused by the gravitational tug of stellar companions \citep{Perryman12}. 
While the impact of astrometry on exoplanet detection has so far been limited, 
the technique has enormous potential and is complementary to other methods 
\citep{Gatewood76, Black82, Sozzetti05}. 
Astrometry is most sensitive to wider orbits, because the center of mass displacement amplitude 
increases with orbital period.  As a result, detectable 
orbital periods are typically several years.  The need for measurement stability and 
precision over such long time baselines has been a challenging requirement for 
currently available instruments. Fortunately, with the successful launch of the Gaia satellite, the prospects for space-based astrometric 
planet searches are good.

\subsubsection{Parametrization of orbital motion} 
\begin{figure}[!htb]
\includegraphics[width= \columnwidth]{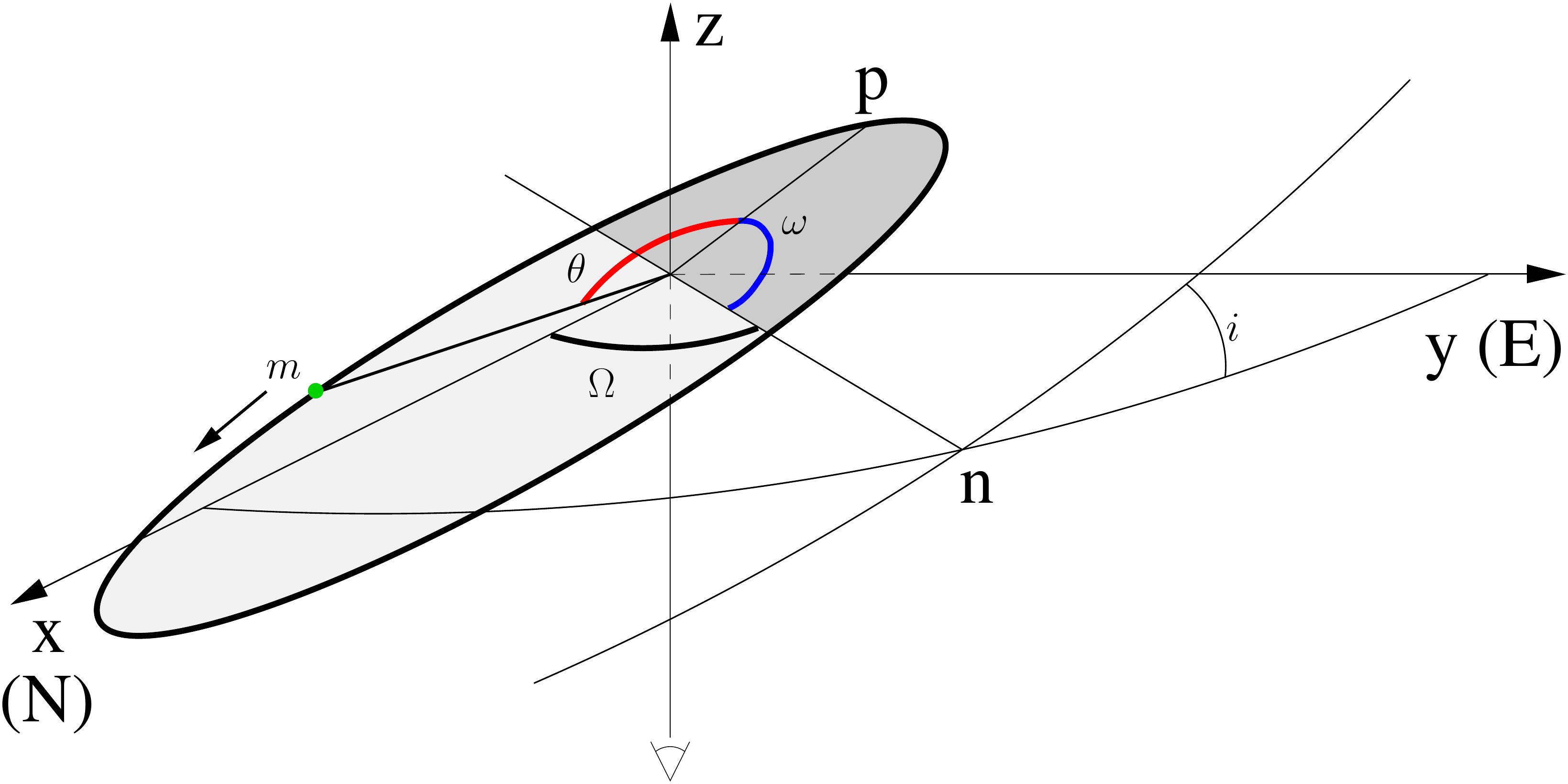}%
\caption{Illustration of the orbit described by a star ($m$) about the barycenter located at the origin. The observer sees the sky plane defined by the x-y axes from below along the z axis. The angles $i$, $\omega$, $\Omega$, $\Theta$, the ascending node n, and the periastron position p are indicated. By convention, x is North and y is East. Figure from \cite{Sahlmann12}.}
\label{fig:hilditch}
\end{figure}

The term \emph{astrometry} refers to the measurement of a star's position relative 
to the background sky, i.e.,\ an astrometric orbit corresponds to the barycentric motion 
of a star caused by an invisible companion. This motion follows Kepler's laws and is 
parametrized by the period $P$, the eccentricity $e$, the time of periastron passage $T_0$, its inclination relative to the sky plane $i$, the longitude of periastron $\omega$, the longitude of the ascending node $\Omega$, and the semi-major axis $a_1$ expressed in angular units (Fig.~\ref{fig:hilditch}). The Thiele-Innes constants $A,B,F,G$ are commonly used instead of the parameters $a_1$, $\omega$, $\Omega$, $i$, because they linearize the orbit term in the general expression for an astrometric signal $\Lambda$ measured along an axis determined by the angle $\psi$ 
\begin{equation}\label{eq:abscissa1}
\begin{split}
\Lambda = \,& ( \Delta \alpha^{\star} + \mu_{\alpha^\star} \, t ) \, \cos \psi + ( \Delta \delta + \mu_\delta \, t ) \, \sin \psi + \varpi \, \Pi_\psi \\
& + (B \, X + G \, Y) \cos \psi + (A \, X + F \, Y) \sin \psi,
\end{split}\end{equation}   
where $\varpi$ is the parallax, $\Pi_\psi$ is the parallax factor along $\psi$, $X$ and $Y$ are the rectangular coordinates \citep{Hilditch01}
\begin{equation}
X = \cos E - e \hspace{5mm} Y = \sqrt{1-e^2} \sin E,
\end{equation}
and $E$ is the eccentric anomaly. This relation includes coordinate offsets in the equatorial system ($\Delta \alpha^{\star},\Delta \delta$), proper motions ($\mu_{\alpha^\star},\mu_\delta$), parallactic motion, and orbital motion. It can be applied to both one and two-dimensional measurements made by Hipparcos, Gaia, or interferometers.

\subsubsection{Signal dependence on mass and distance}
\begin{figure}[!htb]
\begin{center}
\includegraphics[width = \columnwidth]{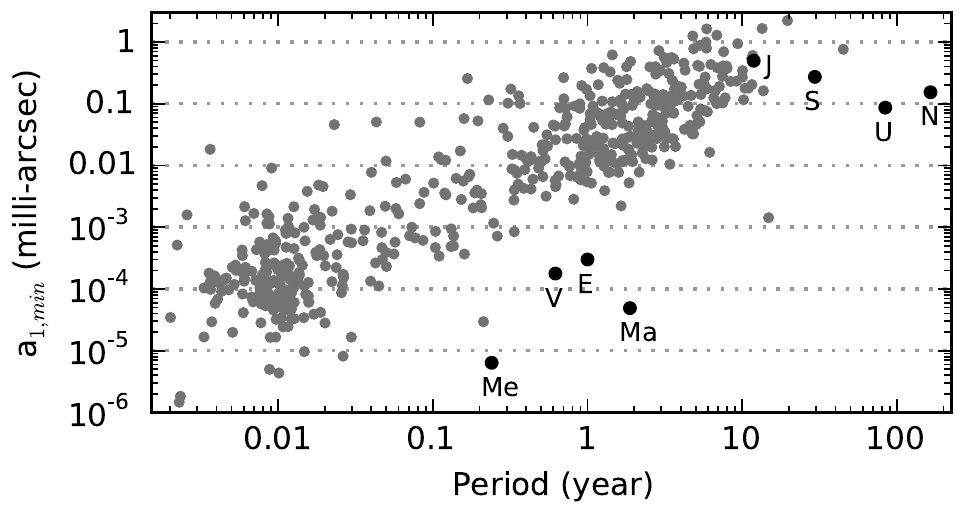}
\caption{Minimum astrometric signature of the host star as a function of orbital period for 570 planets (grey circles). For reference, the astrometric signatures of a solar-mass star located at a distance of 10 pc caused by the solar system planets are shown with black circles and labelled with the planet initials.}
\label{fig:planetsignatures}\end{center}\end{figure}

The semi-major axis $\bar a_{1}$ of a star's barycentric orbit is related to the stellar mass $m_1$, the mass of the companion $m_2$, and the orbital period by Kepler's third law (SI units)
\begin{equation}\label{eq:kthird}
 4 \pi^2\; \frac{\bar a_{1}^3}{P^2} = G\;\frac{M_P^3}{(M_*+M_P)^2},
\end{equation}
where $G$ is the gravitational constant. The relation between angular and linear semi-major axes is proportional to the parallax, $a_1 \propto \varpi \,\bar a_1$, thus the orbit's apparent angular size decreases reciprocally with the system's distance from Earth. The value of $a_1$ determines the semi-amplitude of the periodic signal we intend to detect with astrometric measurements. Figure~\ref{fig:planetsignatures} shows the minimum astrometric signature $a_{1,\mathrm{min}}$ derived from Eq.~\ref{eq:kthird} for planets listed in the exoplanets.org database \citep{Wright11} on June 1, 2013, that have an entry for distance, star mass, orbital period, and planet mass, where we assumed circular orbits. For radial velocity planets, $a_{1,\mathrm{min}}$ is a lower bound because we set $\sin i=1$. The spread at a given period originates in differing distances, star masses, and planet masses. Figure~\ref{fig:planetsignatures} illustrates the typical signal amplitudes for the known exoplanet population and highlights that only a small fraction of known planets is accessible with a measurement precision of 1 milli-arcsec (mas). It also shows that an improvement by only one order of magnitude in precision would set astrometry over the threshold of routine exoplanet detection.

\subsubsection{Scientific potential} 
The motivation for using astrometry to carry out exoplanet searches is founded in the rich and complementary orbital information provided by this technique. Astrometric measurements determine the value of $m_2^3/(m_1\!+m_2)^2$, thus if the host star mass is known, then planet mass $m_2$ can be estimated without the $\sin i$ ambiguity of radial velocity measurements. An astrometric study of a statistical sample of exoplanets could therefore accurately determine the planet mass function and help to refine theories of planet formation. Equation~\ref{eq:kthird} implies that any orbital configuration creates an astrometric signal and the amplitude increases with orbital period (see the trend in Fig.~\ref{fig:planetsignatures}), making astrometry an ideal technique for the study of planets on long-period orbits. Because the technique measures the photocenter, it is sensitive to the detection of planets around fast rotating stars with broad spectral lines or around very faint objects like brown dwarfs. There may also be a reduced sensitivity to stellar activity compared to radial velocity or photometric measurements \citep{Eriksson07, Lagrange11}. Since activity is currently hampering the detection of Earth-mass planets (e.g.,\ \citealt{Dumusque12}), astrometry may hold a distinct advantage for future searches, although the precision needed to detect Earth-like planets around the closest stars is at the level of 1 micro-arcsecond. Astrometry is applicable to planet searches around nearby stars of various masses and ages, with benefits for the study of the planet mass function, of long-period planets, and of planets around active stars.

\subsection{Techniques and Instruments}

The precision $\sigma$ of an astrometric measurement is fundamentally limited by the ability to measure an image position on a detector. In the diffraction limit, it is therefore related to the wavelength $\lambda$, the aperture size $D$, and the signal-to-noise S/N, typically limited by photon noise $\mathrm{S/N}\!\sim\!\sqrt{N_p}$
\begin{equation}
\sigma \propto \frac{1}{\mathrm{S/N}}\frac{\lambda}{D},
\end{equation}
thus, the achievable astrometric precision improves with the aperture size. For observations from the ground, the turbulence in the Earth's atmosphere above the telescope is the dominant error source. It can be mitigated by modeling of seeing-limited observations \citep{Lazorenko04}, by the use of adaptive optics \citep{Cameron09}, and with off-axis fringe tracking in dual-field interferometry \citep{Shao92}. Space-borne instruments avoid atmospheric perturbations altogether and give access to nearly diffraction-limited observations, thus are ideal for high-precision astrometry work. Regardless of how the data were collected, the number of free astrometric parameters of a system with $n$ planets is $5+n\times7$, i.e., at least 12 (see Eq.~\ref{eq:abscissa1}), compared to $1+n\times5$ parameters for a radial velocity orbit adjustment \citep{Wright09b}. To obtain a robust solution and to minimize parameter correlations, for instance between proper, parallactic, and orbital motion, a minimum timespan of one year and appropriate sampling of the orbital period are required. 

\subsubsection{Ground-based astrometry}
Repeated imaging of a target and the measurement of its motion relative to background sources is a basic astrometric method and several planet search surveys use seeing-limited optical imaging with intermediate and large telescopes \citep{Pravdo96, Boss09}. Accuracies of better than 0.1 mas have been achieved with this method \citep{Lazorenko09}, which satisfies the performance improvement necessary for efficient planet detection. Adaptive-optics assisted imaging is also being used, for instance for a planet search targeting binaries with separations of a few arcseconds \citep{Roll11}. An optical interferometer realizes a large effective aperture size by combining the light of multiple telescopes  that translates into an achievable precision of 0.01 milliarcseconds in the relative separation measurement of two stars typically less than 1\arcmin~apart \citep{Shao92}. Several observatories have implemented the necessary infrastructure and are pursuing astrometric planet search programmes \citep{Launhardt08, Muterspaugh10, Woillez10, Sahlmann13a}. Similarly, Very Long Baseline Radio Interferometry is a promising method for targeting nearby stars sufficiently bright at radio wavelengths. \citep{Bower09}.

\subsubsection{Astrometry from space}
Space astrometry was firmly established by the Hipparcos mission that operated in 1989-1992 and resulted in the determination of positions, proper motions, and absolute parallaxes at the 1~mas level for $120\,000$ stars \citep{Perryman97}. The satellite's telescope had a diameter of only 29~cm and scanned the entire celestial sphere several times to construct a global and absolute reference frame. However, Hipparcos data do not have the necessary precision to determine the astrometric orbits of the majority of know exoplanets. On a smaller scale but with slightly better precision, the Hubble space telescope fine guidance sensor has made  stellar parallax and orbit measurements possible \citep{Benedict01}. Because the Stellar Interferometry Mission \citep{Unwin08} was discontinued, Gaia is the next space astrometry mission capable of detecting extrasolar planets. 

\subsection{Results from Astrometry}
\subsubsection{Combination with radial velocities}
\begin{figure}[!htb]
\includegraphics[width= \columnwidth]{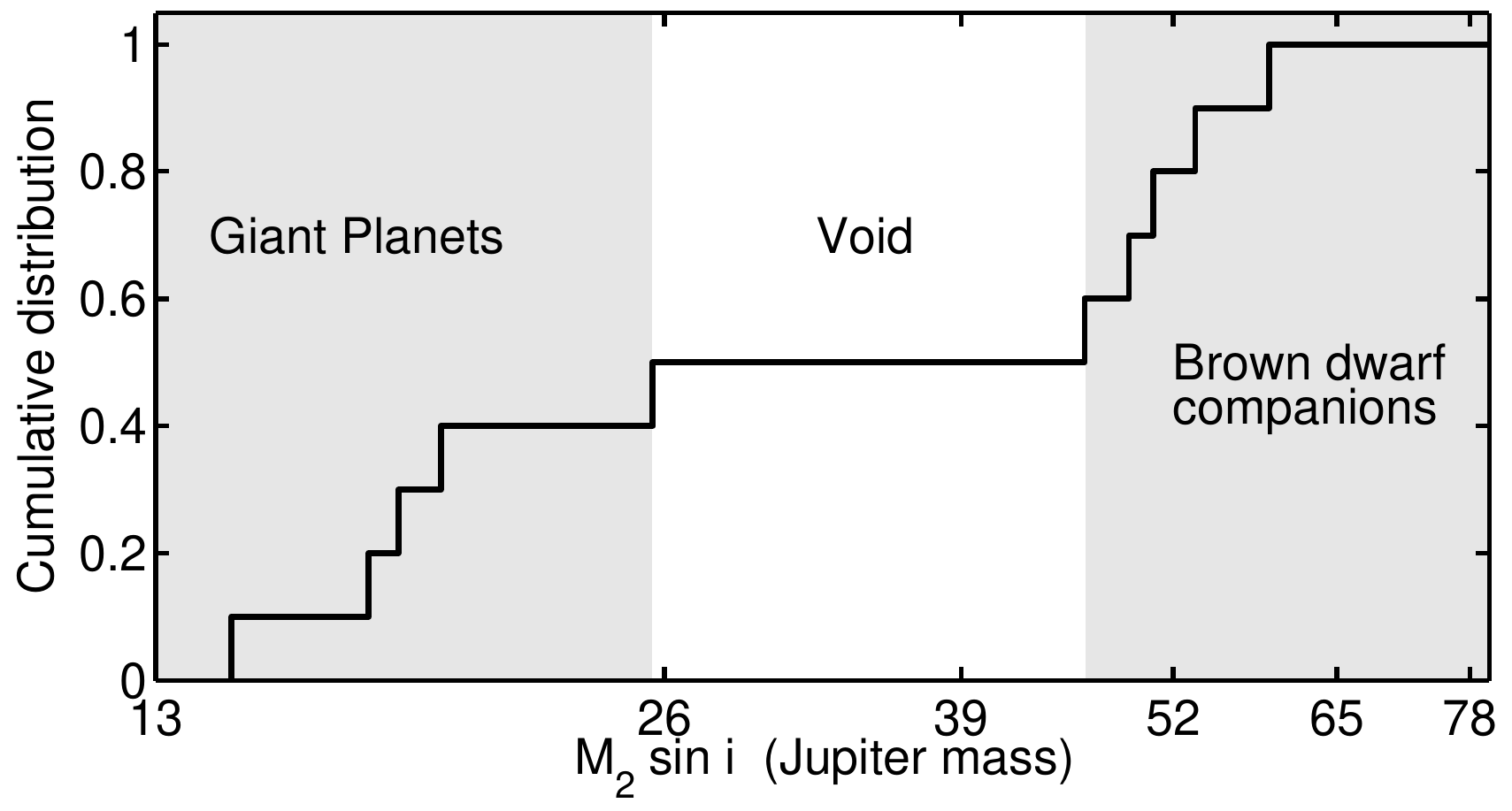}
\caption{The minimum mass distribution of substellar companions within 10 AU of Sun-like stars from the Coralie RV survey after constraining the orbital inclinations with Hipparcos astrometry.}
\label{fig:dividingline}
\end{figure}

For a planet detected with radial velocities (RV), five out of seven orbital parameters are constrained. The two remaining parameters, the inclination $i$ and $\Omega$, can be determined by measuring the astrometric orbit. The knowledge of the RV parameters (or the high weight of RV measurements) leads to a significant reduction of the required S/N for a robust astrometric detection. Second, even an astrometric non-detection carries valuable information, e.g.,\ an upper limit to the companion mass. Therefore, this type of combined analysis is so far the most successful application of astrometry in the exoplanet domain. Hipparcos astrometry yielded mass upper limits of RV planets \citep{Perryman96, Torres07, Reffert11} and revealed that, in rare cases, brown dwarf \citep{Sahlmann11a} or stellar companions \citep{Zucker01} are mistaken for RV planets because their orbital planes are seen with small inclinations. Similarly, the Hubble fine guidance sensor was used to determine the orbits and masses of brown dwarf companions to Sun-like stars initially detected with RV \citep{Martioli10, Benedict10} and ground-based imaging astrometry yielded a mass upper limit of $\sim 3.6\,M_J$ to the planet around GJ317 \citep{Anglada-E2012}. \citet{Sahlmann11b} (Fig.~\ref{fig:dividingline}) used Hipparcos data to eliminate low-inclination binary systems mimicking brown dwarf companions detected in a large RV survey, revealing a mass range where giant planets and close brown dwarf companions around Sun-like stars are extremely rare. 

\subsubsection{Independent discoveries}
\begin{figure}[!htb]
\center
\includegraphics[width= 0.95\columnwidth]{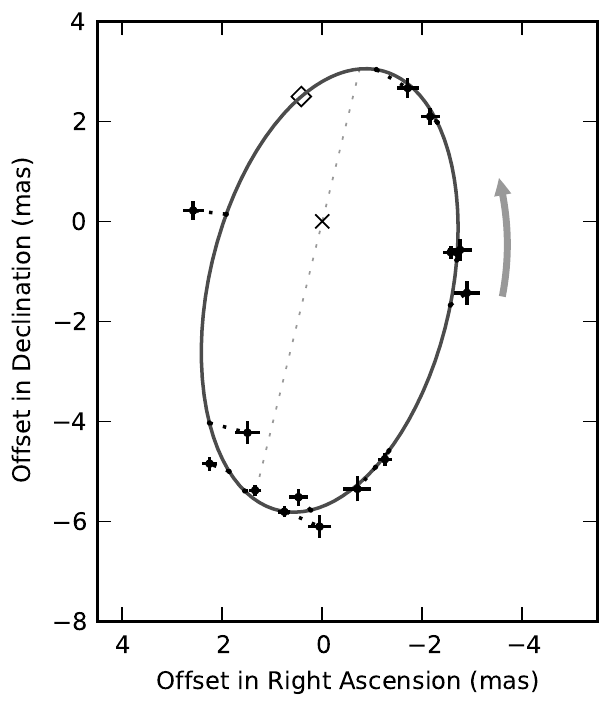}
\caption{The barycentric orbit of the L1.5 dwarf {DENIS-P\,J082303.1-491201} caused by a 28 Jupiter mass companion in a 246 day orbit discovered through ground-based astrometry with an optical camera on an 8 m telescope \citep{Sahlmann13b}. }
\label{fig:bdo}
\end{figure}

Working towards the goal of exoplanet detection, optical imaging surveys have succeeded in measuring the orbits of low-mass binaries and substellar companions to M dwarfs \citep{Pravdo05, Dahn08}, relying on astrometric measurements only. Interferometric observations revealed the signature of a Jupiter-mass planet around a star in an unresolved binary \citep{Muterspaugh10}, which, if confirmed independently, represents the first planet discovered by astrometry. Recent improvements of imaging astrometry techniques towards 0.1 mas precision made the discovery of a 28 $M_J$ companion to an early L dwarf possible (Fig.~\ref{fig:bdo}) and demonstrated that such performance can be realised with a single-dish telescope from the ground.

\subsection{The Future}

Without a doubt, our expectations are high for the Gaia mission which was launched on 19 December 2013. Gaia is a cornerstone mission of the European Space Agency that will implement an all-sky survey of an estimated billion stellar objects with visible magnitudes of 6--20 \citep{Perryman01, deBruijne12}. On average, the astrometry of a star will be measured 70 times over the mission lifetime of five years with a single measurement precision of $\sim$0.02-0.05\,mas for stars brighter than $\sim$14th magnitude. Another look at Fig.~\ref{fig:planetsignatures} shows that hundreds of known exoplanet systems will be detectable and it is expected that Gaia will discover thousands of new exoplanets \citep{Casertano08}, yielding a complete census of giant exoplanets in intermediate-period orbits around nearby stars. The sight of astrometric orbits caused by planets around stars will then become just as common as radial velocity curves and dips in light-curves are today. Assuming that it will perform as planned, Gaia will therefore add astrometry to the suite of efficient techniques for the study of exoplanet populations and will help us to advance our understanding of (exo-)planet formation. It will also pave the way for future space astrometry missions aiming at detecting the Earth-like planets around nearby stars \citep{Malbet12}.\\  

At the same time, ground-based surveys will remain competitive because they offer long lifetimes, scheduling flexibility, and access to targets not observable otherwise. They are also necessary for technology development and demonstration. The upcoming generation of sub-mm/optical interferometers and telescopes will have larger apertures and wide-field image correction, and hence provide us with even better astrometric performance and new opportunities for exoplanet science.

\section{\textbf{Statistical Distributions of Exoplanet Properties}}

\begin{figure*}[!htb]
\includegraphics[width=0.98\linewidth]{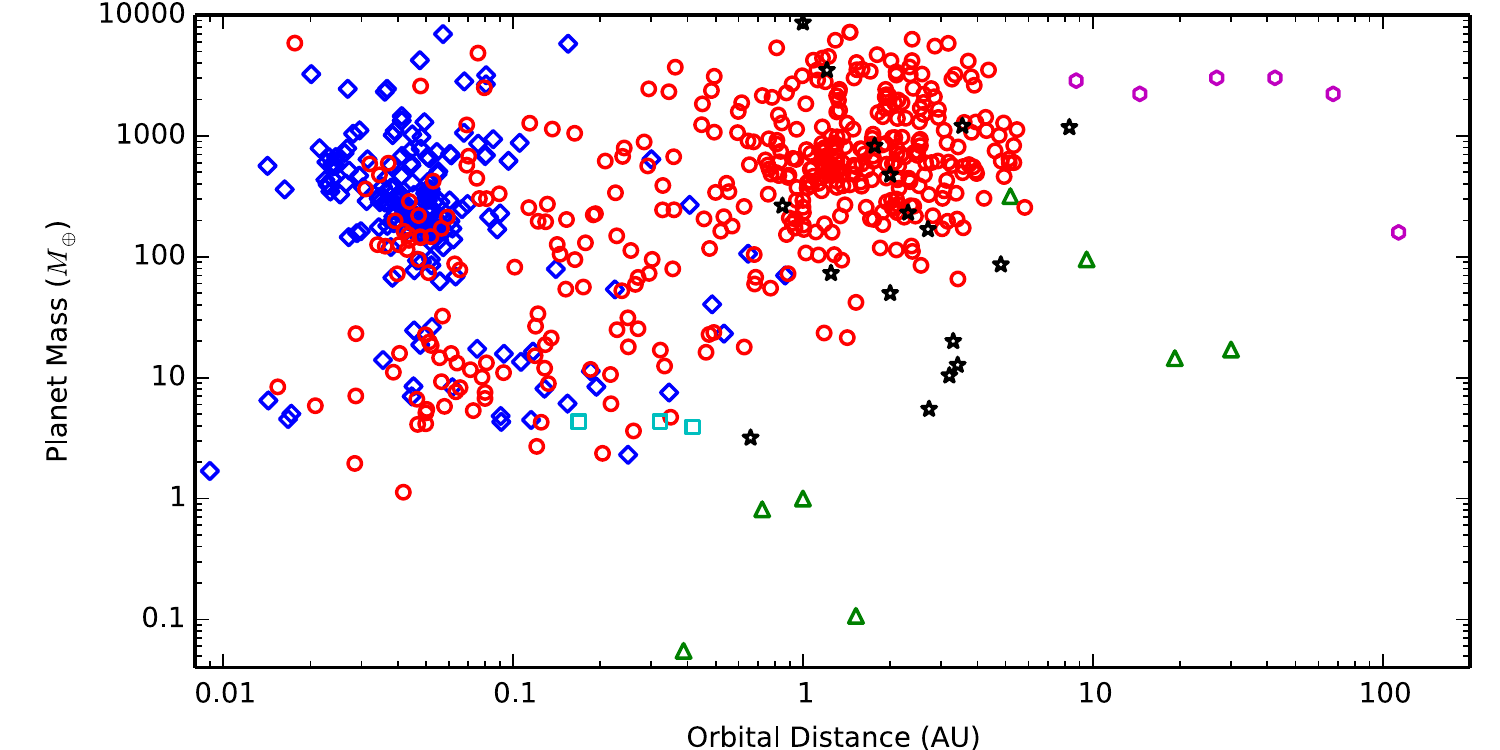}
\caption{Masses and orbital distances of planets from the Exoplanet Orbit Database \citep[][http://exoplanets.org]{Wright11} as of July, 2013.  
The recently discovered Earth-size planet, Kepler-78b, is also included \citep{Sanchis-Ojeda2013,Howard2013b,Pepe2013}.
Extrasolar planets are color-coded according to their method of discovery: RV = red circles, transit = blue diamonds, imaging = magenta hexagons, gravitational microlensing = black stars, and pulsar timing = cyan squares.  
Planets in the Solar System are green triangles.  
Projected semi-major axis is plotted for microlensing planets while true semi-major axis is plotted for others. 
The occurrence of some planet types (e.g., hot Jupiters) are exaggerated relative to their true occurrence 
due to their relative ease of discovery.}
\label{fig:mass_asemi}
\end{figure*}

In this section, we review and interpret the major statistical properties of extrasolar planets. 
We focus primarily on results from RV and transit surveys since they have produced the bulk of the discovered planets.  
Fig.\ \ref{fig:mass_asemi} shows known planets with measured masses and semi-major axes (projected for microlensing planets).  
The major archetypes of well-studied  planets---cool Jupiters in $\sim$1--5 AU orbits, hot Jupiters in sub-0.1 AU orbits, and sub-Neptune-size planets 
orbiting within 1 AU---are all represented, although their relative frequencies are exaggerated due to differing survey sizes and yields.
For more thorough reviews of exoplanet properties, the reader is directed to the literature \citep{Howard2013,Cumming2011,Marcy2005,Udry2007}.

\subsection{Abundant, close-in small planets}

\begin{figure}[!htb]
\includegraphics[width= \columnwidth]{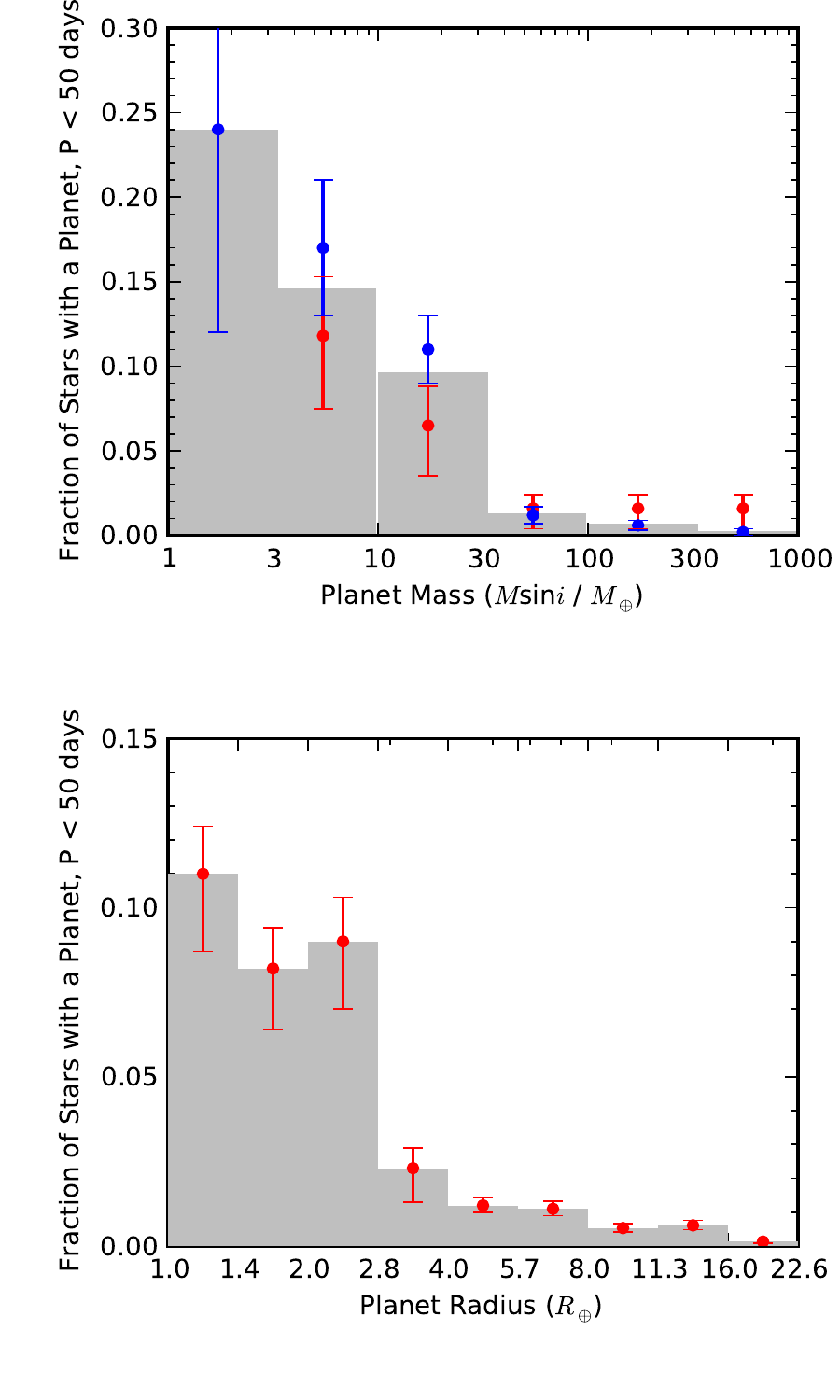}
\caption{The mass (top) and size (bottom) distributions of planets orbiting close to G and K-type stars.  The distributions rise substantially with decreasing size and mass, indicating that small planets are more common than large ones.  Planets smaller than 2.8 \rearth\ or less massive than 30 \mearth are found within 0.25 AU of 30--50\% of Sun-like stars. (A) The size distribution is drawn from two studies of Kepler data: \citet{Petigura2013} for planets smaller than four times Earth size and \citet{Howard2012} for larger planets. The mass (\msini) distributions show the fraction of stars having at least one planet with an orbital period shorter than 50 days (orbiting inside of $\sim$0.25 AU) are from separate Doppler surveys (red = \cite{Howard2010}, blue = \cite{Mayor2011}), while the histogram shows their average values.  Both distributions are corrected for survey incompleteness for small/low-mass planets to show the true occurrence of planets in nature.}
\label{fig:mass_radius_histo}
\end{figure}

Planets intermediate in size between Earth and Neptune are surprisingly common in extrasolar systems, 
but notably absent in our Solar System. 
The planet size and mass distributions (Fig. \ref{fig:mass_radius_histo}) demonstate that small planets substantially outnumber 
large ones, at least for close-in orbits.  
Doppler surveys using  HIRES  at Keck Observatory \citep{Howard2010} 
and HARPS \citep{Lovis2009,Mayor2011} at the 3.6 m ESO telescope have 
shown that small planets (Neptune size and smaller) significantly outnumber large ones 
for close-in orbits.  
Using the detected planets and detection completeness contours, 
the Eta-Earth Survey at Keck found that the probability of a star hosting a close-in planet scales as $(M \sin i)^{-0.48}$: small planets are more common.  
In absolute terms, 15\% of Sun-like stars host one or more planets with \msini = 3--30 \mearth\ orbiting within 0.25 AU.
The HARPS survey  confirmed the rising planet mass function with decreasing mass and extended it to 1--3 \mearth planets.  It also demonstrated that low-mass planets have small orbital eccentricities and are commonly found in multi-planet systems with 2--4 small planets orbiting the same star with orbital periods of weeks or months.  It found that at least 50\% of stars have one or more planets of any mass with $P$ $<$ 100 days.

The distribution of planet sizes (radii) measured by the Kepler mission (Fig. \ref{fig:mass_radius_histo}) 
follows the same qualitative trend as the mass distribution, with small planets being more common \citep{Howard2013,Petigura2013,Fressin2013}.  
However, the planet radius distribution extends with small error bars down to 1 \rearth\ for close-in planets, 
while the mass distribution has 50\% uncertainty level near 1 \mearth.  The size distribution is characterized by a power-law rise in occurrence with decreasing size \citep{Howard2013} down to a critical size of $\sim$2.8 \rearth, below which planet occurrence plateaus \citep{Petigura2013}.  The small planets detected by Kepler ($<$2 \rearth) appear to have more circular orbits than larger planets \citep{Plavchan2012}, suggesting reduced dynamical interactions.

The high occurrence of small planets with $P$ $<$ 50 days likely extends to more distant orbits.  As Kepler accumulates photometric data, it becomes sensitive to planets with smaller sizes and longer orbital periods.  Based on 1.5 years of photometry, the small planet occurrence distribution as a function of orbital period is flat to $P$ = 250 days (with higher uncertainty for larger $P$).  Quantitatively, the mean number of planets per star per logarithmic period interval is proportional to $P^{+0.11\pm0.05}$ and $P^{-0.10\pm0.12}$ for 1--2 \rearth and 2--4 \rearth planets, respectively \citep{Dong2012}.

The Kepler planet distribution also shows that small planets are more abundant around around cool stars \citep[][although see Fressin et al. 2013 for an opposing view]{Howard2012}.  M dwarfs observed by Kepler appear to have a high rate of overall planet occurrence, 0.9 planets per star in the size range 0.5--4 \rearth\ in $P$ $<$ 50 day orbits.  Earth-size planets (0.5-1.4 \rearth) are estimated to orbit in the habitable zones (HZ) of $15^{+13}_{-6}$\% of Kepler's M dwarfs \citep{Dressing2013}.  This estimate depends critically on the orbital bounds of the habitable zone; using more recent HZ models, the fraction of M dwarfs with Earth-size planets in the HZ may be three times higher \citep{Kopparapu2013}.

\begin{figure}[!htb]
\includegraphics[width= \columnwidth]{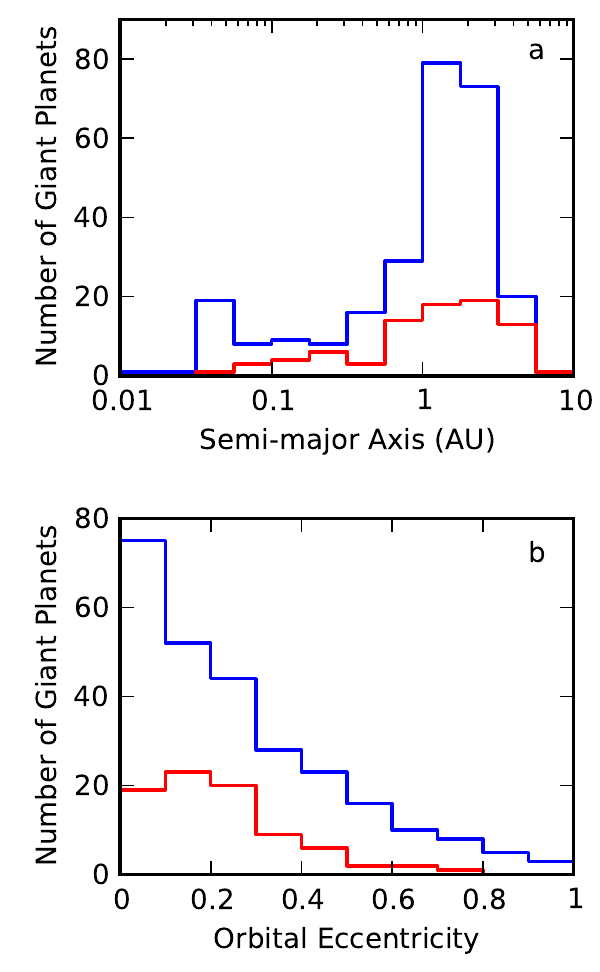}
\caption{Orbital characteristics of giant planets ($M_P \sin i$ $>$ 0.2 $M_J$) detected by Doppler surveys as cataloged on the Exoplanet Orbit Database \citep{Wright11}.  The number distribution of semi-major axes (a) shows that apparently single planets (blue) preferentially orbit at distances of $\sim$0.05 AU and at $\sim$1-3 AU from their host stars.  These preferred orbits are diminished in multi-planet systems (red).  The decline in number of detected planets for orbits outside of $\sim$3 AU is not significant; fewer stars have been searched for such planets compared to the closer orbits.  The distribution of orbital eccentricities (b) for apparently single planets (blue) span the full range, with low-eccentricity orbits being more common.  Giant planets in multi-planet systems (red) have orbits that are more commonly close to circular.  The larger eccentricities of single planets suggests that they were dynamically excited from a quiescent, nearly circular origin, perhaps by planet-planet scattering that resulted in the ejection of all but one detectable planet per system.}
\label{fig:gas_giants}
\end{figure}

Of the Kepler planet host stars, 23\% show evidence for two or more transiting planets.  To be detected, planets in multi-transiting systems likely orbit in nearly the same plane, with mutual inclinations of a few degrees at most.  The true number of planets per star (transiting or not) and their mutual inclinations can be estimated from simulated observations  constrained by the number of single, double, triple, etc. transiting systems detected by Kepler \citep{Lissauer2011b}.  \citet{Fang2012} find an intrinsic multi-planet distribution with 54\%, 27\%, 13\%, 5\%, and 2\% of systems having 1, 2, 3, 4, and 5 planets with $P$ $<$ 200 days.  Nearly all multi-planet systems (85\%) have mutual inclinations of less than 3$^\circ$ \citep{Fang2013, Johansen2012}.  Mutual inclinations of a few degrees are also suggested by comparison between the Kepler and HARPS data \citep{Figueira2012}.  This high degree of co-planarity is consistent with planets forming in a protoplanetary disk without significant dynamical perturbations.  

The ratios of orbital periods in multi-transiting systems provide additional dynamical constraints.  These ratios are largely random \citep{Fabrycky2012}, with a modest excess just outside of period ratios that are consistent with dynamical resonances (ratios of 2:1, 3:2, etc.) and a compensating deficit inside \citep{Lithwick2012}.  The period ratios of adjacent planet pairs demonstrate that $>$31, $>$35, and $>$45\% of 2-planet, 3-planet, and 4-planet systems are dynamically packed; adding a hypothetical planet would gravitationally perturb the system into instability \citep{Fang2013}.  

\subsection{Gas giant planets}

The orbits of giant planets are the easiest to detect using the Doppler technique and were the first to be studied statistically \citep[e.g.][]{Udry2003,Marcy2005}.  Observations over a decade of a volume-limited sample of $\sim$1000 F, G, and K-type dwarf stars at Keck Observatory showed that 10.5\% of G and K-type dwarf stars host one or more giant planets (0.3--10 $M_J$) with orbital periods of 2--2000 days (orbital distances of $\sim$0.03--3 AU).  Within those parameter ranges, less massive and more distant giant planets are more common.  
Extrapolation of this model suggests that 17--20\% of such stars have giant planets orbiting within 20 AU ($P$ = 90 years) \citep{Cumming08}.  This extrapolation is consistent with a measurement of giant planet occurrence beyond $\sim$2 AU from microlensing surveys \citep{Gould10}.  However, the relatively few planet detections from direct imaging planet searches suggest that the extrapolation is not valid beyond $\sim$65 AU \citep{Nielsen2010}.

These smooth trends in giant planet occurrence mask  pile-ups in semi-major axis. \citep{Wright2009a}.  The  orbital distances for giant planets show a preference for orbits larger than $\sim$1 AU and to a lesser extent near 0.05 AU (``hot Jupiters'') (Fig.\ \ref{fig:gas_giants}a).  This �period valley� for apparently single planets is interpreted as a transition region between two categories of Jovian planets with different migration histories \citep{Udry2003}.  The excess of planets starting at $\sim$1 AU approximately coincides with the location of the ice line, which provides additional solids that may speed the formation of planet cores or act as a migration trap for planets formed farther out \citep{Ida2008}.  The semi-major axis distribution for giant planets in multi-planet systems is more uniform, with hot Jupiters nearly absent and a suppressed peak of planets in $>$1 AU orbits.  

The giant planet eccentricity distribution (Fig.\ \ref{fig:gas_giants}b) also differs between single and multi-planet systems.  The eccentricities of single planets can be reproduced  by a dynamical model in which initially low eccentricities are excited by planet-planet scattering \citep{Chatterjee2008}.  Multi-planet systems with a giant planet likely experienced substantially fewer scattering events.  
The single planet systems may represent the survivors of scattering events that ejected other planets in the system.

\begin{figure}[!htb]
\center
\includegraphics[width= 0.9\columnwidth]{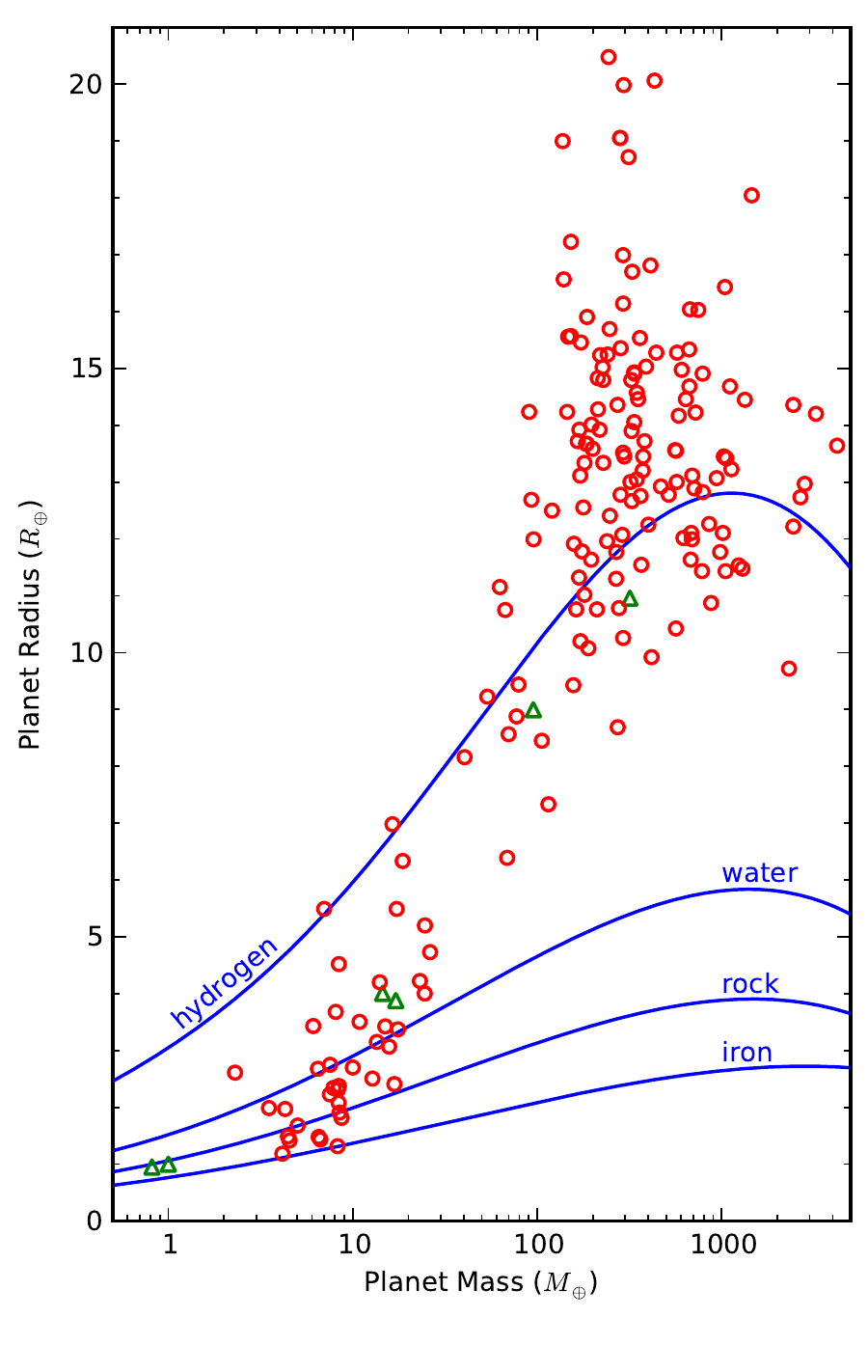}
\caption{Masses and radii of well-characterized planets from the Exoplanet Orbit Database 
\citep{Wright11}. Extrasolar planets are shown as open red circles and  solar system planets are open green triangles. 
Blue lines show model mass-radius relationships for idealized planets consisting of 
pure hydrogen \citep{Seager2007}, water, rock (Mg$_2$SiO$_4$), or iron \citep{Fortney2007}. 
Poorly understood heating mechanisms inflate some gas giant planets 
(larger than $\sim$8 $R_E$) to sizes larger than predicted by the simple hydrogen model. 
Smaller planets (less massive than $\sim$30 $M_E$) show great diversity in size at a fixed mass, 
likely due to varying density of solids and atmospheric extent. }
\label{fig:mass_radius}
\end{figure}

Metal-rich stars are more likely to host giant planets within 5 AU.  This ``planet-metallicity correlation'' was validated statistically by Doppler surveys of  stars with $M_{\star}$ = 0.7--1.2 $M_{\sun}$ and uniformly-measured metallicities \citep{Fischer2005,Santos2004}.  The probability of a star hosting a giant planet is proportional to the square of the number of iron atoms in the star relative to the Sun, ${\cal P}$(planet) $\propto$ N$_{\rm Fe}^2$.  A later Doppler study spanned a wider range of stellar masses (0.3--2.0 $M_{\sun}$) and showed that the probability of a star hosting a giant planet correlates with both stellar metal content and stellar mass, ${\cal P}$(planet) $\propto$ N$_{\rm Fe}^{1.2\pm0.2} M_{\star}^{1.0\pm0.3}$ \citep{Johnson10b}.  Note that the planet-metallicity correlation only applies to gas giant planets.  Planets larger than 4 \rearth\ (Neptune size) preferentially orbit metal-rich stars, while smaller planets are are non-discriminating in stellar metallicity \citep{Buchhave2012}.  This pattern of host star metallicity can be explained if small planets  commonly form in protoplanetary disks, but only a fraction of those small planets grow to a critical size in time to become gas giants.

Although hot Jupiters (giant planets with $P$  $\lesssim$ 10 days) are found around only 0.5-1.0\% of Sun-like stars \citep{Wright2012}, they are the most well-characterized planets because they are easy to detect and follow up with ground- and space-based telescopes.  However, their origin remains mysterious.  In contrast to the commonly multiple sub-Neptune-size planets, hot Jupiters are usually the only detected planet orbiting the host star within observational limits \citep{Steffen2012}.  Many hot Jupiters have low eccentricities due tidal circularization.  The measured obliquities of stars hosting hot Jupiters display a peculiar pattern: obliquities are apparently random above a critical stellar temperature of $\sim$6250 K, but cooler systems are mostly aligned.  In situ formation is unlikely for hot Jupiters because of insufficient protoplanetary disk mass so close to the star.  It is more likely that they formed at several AU, were gravitationally perturbed into orbits with random inclinations and high eccentricities, and were captured at $\sim$0.05 AU by dissipation of orbital energy in tides raised on the planet.  For systems with sufficiently strong tides raised by the planet on the star (which depend on a stellar convective zone that is only present below for $T_{\rm eff} \lesssim$ 6250 K), the stellar spin axis aligns to the orbital axis (12).

\subsection{Mass-radius relationships}

While the mass and size distributions provide valuable information about the relative occurrence of planets of different types, it remains challenging to connect the two.  Knowing the mass of a planet only weakly specifies its size, and vice versa.  
This degeneracy can be lifted for $\sim$200 planets with well-measured masses and radii (Fig.\ \ref{fig:mass_radius}), most of which are transiting hot Jupiters.  The cloud of points follows a diagonal band from low-mass/small-size to high-mass/large-size.  This band of allowable planet mass/size combinations has considerable breadth.  Planets less massive than $\sim$30 \mearth\ vary in size by a factor of $\sim$5 and planets larger than $\sim$100 \mearth\ (gas giants) vary by a factor of $\sim$2.  For the gas giants, the size dispersion at a given mass is due largely to two effects.  First, planets in tight orbits receive higher stellar flux and are more commonly inflated.  While higher stellar flux correlates with giant planet inflation \citep{Weiss2013}, it is unclear how the stellar energy is deposited in the planet�s interior.  Less importantly, the presence of a massive solid core (or distributed heavy elements) increases a planet�s surface gravity, causing it be more compact.  

Low-mass planets show an even larger variation in size and composition.  Three examples of sub-Neptune-size planets illustrate the diversity.  The planet Kepler-10b has a mass of 4.6 \mearth\ and a density of 9 g cm$^{-3}$, indicating a rock/iron composition and no atmosphere \citep{Batalha2011}.  In contrast, the planet Kepler-11e has a density of 0.5 g cm$^{-3}$ and a mass of 8 \mearth.  A substantial light-element atmosphere (probably hydrogen) is required to explain its mass and radius combination \citep{Lissauer2011}.  The masses and radii of intermediate planets lead to ambiguous conclusions about composition.  For example, the bulk physical properties of GJ 1214b \citep[6.5 \mearth, 2.7 \rearth,  1.9 g cm$^{-3}$,][]{Charbonneau2009} are consistent with several compositions: a ``super-Earth'' with a rock/iron core surrounded by $\sim$3\% H$_2$ gas by mass; a �water world� planet consisting of a rock/iron core, a water ocean and atmosphere that contribute $\sim$50\% of the mass; or a �mini-Neptune� composed of rock/iron, water, and H/He gas \citep{Rogers2010}.

\bigskip

\bibliographystyle{ppvi_lim1}
\bibliography{ppvi_exoplanets}

\end{document}